\documentclass[prd,twocolumn,eqsecnum]{revtex4}
\usepackage{amssymb}
\usepackage{bm}
\begin{document}
\title{Gravitational perturbations of the Schwarzschild spacetime: A 
practical covariant and gauge-invariant formalism} 
\author{Karl Martel and Eric Poisson}
\affiliation{Department of Physics, University of Guelph, Guelph,
Ontario, Canada N1G 2W1}
\date{January 23, 2004} 
\begin{abstract}
We present a formalism to study the metric perturbations of the
Schwarzschild spacetime. The formalism is gauge invariant, and it is
also covariant under two-dimensional coordinate transformations that
leave the angular coordinates unchanged. The formalism is applied to
the typical problem of calculating the gravitational waves produced by
material sources moving in the Schwarzschild spacetime. We examine the
radiation escaping to future null infinity as well as the radiation
crossing the event horizon. The waveforms, the energy radiated, and
the angular-momentum radiated can all be expressed in terms of two
gauge-invariant scalar functions that satisfy one-dimensional wave
equations. The first is the Zerilli-Moncrief function, which satisfies
the Zerilli equation, and which represents the even-parity sector of
the perturbation. The second is the Cunningham-Price-Moncrief
function, which satisfies the Regge-Wheeler equation, and which
represents the odd-parity sector of the perturbation. The covariant
forms of these wave equations are presented here, complete with
covariant source terms that are derived from the stress-energy tensor
of the matter responsible for the perturbation. Our presentation of
the formalism is concluded with a separate examination of the monopole
and dipole components of the metric perturbation.     
\end{abstract} 
\maketitle

\section{Introduction}
\label{i} 

Metric perturbations of the Schwarzschild spacetime have been studied
for a long time, starting with the pioneering work of Regge and
Wheeler \cite{regge-wheeler:57}, Vishveshwara \cite{vishveshwara:70},
and Zerilli \cite{zerilli:70}. The theory was summarized in an
influential monograph by Chandrasekhar \cite{chandrasekhar:83}, and
it has been applied to many different physical situations (see, for
example, Chapter 4 of the book by Frolov and Novikov
\cite{frolov-novikov:98}). In particular, a useful application  
has been the computation of gravitational waves produced by a point 
particle moving in the field of a Schwarzschild black hole (see
Ref.~\cite{martel:04}, the review article of
Ref.~\cite{sasaki-tagoshi:03}, and references therein). Another has
been the simulation of a collision of two black holes in a
``close-limit'' approximation (see the review article of
Ref.~\cite{gleiser-etal:00} and references therein).    

Traditionally the perturbation formalism is developed in the standard
Schwarzschild coordinates $(t,r,\theta,\phi)$, and in a standard
choice of gauge known as the ``Regge-Wheeler gauge.'' The tradition 
also makes use of Fourier-transform techniques and presents the
perturbation equations in the frequency domain instead of the time
domain. Moncrief \cite{moncrief:74} was the first to present the
formalism in a gauge-invariant package, recognizing the practical
advantages that gauge invariance provides: While the Regge-Wheeler
gauge is useful for many purposes, it is not useful for others, and
the power to switch from one gauge to another within a gauge-invariant
framework is often required. Another refinement of the formalism was
produced by Gerlach and Sengupta \cite{gerlach-sengupta:80}, who
presented the gauge-invariant perturbation equations in an arbitrary
coordinate system, thus liberating the formalism from the usual
Schwarzschild coordinates and their poor behavior at the event
horizon.   

This program to translate the traditional perturbation formalism into
a covariant, gauge-invariant language was recently revived by Sarbach
and Tiglio \cite{sarbach-tiglio:01}, and with this paper we hope to
deliver its final chapter. We aim to present the formalism of metric 
perturbations of the Schwarzschild spacetime in its most mature and
practical incarnation yet. Our formalism is covariant and gauge
invariant, and it goes beyond the work of Gerlach and Sengupta, and of
Sarbach and Tiglio, by the development of covariant master equations
for the even-parity and odd-parity sectors of the theory, 
{\it complete with covariant source terms} that are derived from the
stress-energy tensor of the matter which is responsible for the
perturbation.       

We have in mind a typical application of the perturbation formalism,
the calculation of gravitational waves produced by material sources
moving in the Schwarzschild spacetime. We are interested in the
radiation that escapes to future null infinity and manifests itself as
waveforms $h_+$ and $h_\times$ that are directly observable to
gravitational-wave detectors, and we are interested in the radiation
that crosses the black-hole horizon. This application illustrates well
the need for a covariant and gauge-invariant formalism, as the two
types of radiation admit different descriptions. The radiation at
future null infinity is best described by casting the perturbation in
an outgoing radiation gauge and expressing it in a retarded coordinate  
system $(u,r,\theta,\phi)$ related to the usual Schwarzschild
coordinates by the transformation $u = t - r - 2M\ln(r/2M - 1)$.  
The radiation at the horizon, on the other hand, is best described by 
adopting an incoming radiation gauge and expressing the perturbation
in an advanced coordinate system $(v,r,\theta,\phi)$ related to the
usual Schwarzschild coordinates by the transformation $v = t + r 
+ 2M\ln(r/2M - 1)$. Our formalism permits the use of any coordinate
system $x^a = (x^0,x^1)$ that can be obtained from the usual
Schwarzschild coordinates $(t,r)$; we do not, however, consider
transformations of the angular coordinates $\theta^A =
(\theta,\phi)$.  

In the formalism developed in this paper, the radiation at future null
infinity and at the horizon are described in terms of scalar and
gauge-invariant master functions, $\Psi^{lm}_{\rm even}(x^a)$ and 
$\Psi^{lm}_{\rm odd}(x^a)$, which can be computed from the metric 
perturbations. These functions are labeled by spherical-harmonic  
indices $l$ and $m$, and by their behavior under a parity
transformation. The function $\Psi^{lm}_{\rm even}(x^a)$ is
constructed from the even-parity perturbations, and it is equal to 
the gauge-invariant function that was first introduced by Moncrief
\cite{moncrief:74}; it is a close cousin to Zerilli's original master 
function \cite{zerilli:70}, and it satisfies a covariant version of
Zerilli's differential equation. The complete source term for this
equation is presented for the first time in this paper. The function 
$\Psi^{lm}_{\rm odd}(x^a)$, on the other hand, is constructed from 
the odd-parity perturbations, and it is equal to the gauge-invariant
function that was first introduced by Cunningham, Price, and Moncrief 
\cite{cunningham-etal:78} and recently revived by Jhingan and Tanaka
\cite{jhingan-tanaka:03}; it is essentially the time integral of the 
original Regge-Wheeler master function \cite{regge-wheeler:57}, and it 
satisfies a covariant version of the Regge-Wheeler equation. The
complete source term for this equation is presented for the first time
in this paper.  

The paper is organized as follows. In Sec.~\ref{ii} we give a
covariant description of the Schwarzschild spacetime and specify 
our notations and conventions. In Sec.~\ref{iii} we introduce the
scalar, vector, and tensor spherical harmonics that are used in the
decomposition of the metric perturbation. In Sec.~\ref{iv} we examine
the even-parity sector of the perturbation, introduce the
Zerilli-Moncrief master function, and derive the one-dimensional wave
equation that it satisfies. In Sec.~\ref{v} we examine the odd-parity
sector of the perturbation, introduce the Cunningham-Price-Moncrief
master function, and derive the one-dimensional wave equation that it
satisfies. In Sec.~\ref{vi} we describe the behavior of the
perturbations near future null infinity, construct the radiative part
of the perturbation field, extract the waveforms $h_+$ and $h_\times$,
and compute the rates at which the radiation carries away energy and
angular momentum. In Sec.~\ref{vii} we describe the behavior of the
perturbations near the black hole's event horizon, and calculate the
rates at which they transfer energy and angular momentum to the black
hole. In Sec.~\ref{viii} we conclude our presentation and examine the 
nonradiating pieces of the metric perturbation, those associated with
the low multipoles $l=0$ and $l=1$.  

Various technical details are relegated to the Appendices. In Appendix
\ref{A} we expand our discussion of vectorial and tensorial spherical
harmonics. In Appendix \ref{B} we present the perturbed Ricci tensor
for a general spherically-symmetric background spacetime. In
Appendices \ref{C}, \ref{D}, and \ref{E} we list the perturbation
equations in the commonly used coordinate systems $x^a = (t,r)$, 
$x^a = (u,r)$, and $x^a = (v,r)$, respectively. 

Because the topic of metric perturbations of the Schwarzschild
spacetime is so venerable, we will allow ourselves in this paper to 
simply state our results and omit most lengthy derivations that lead
to those results. We hope, however, that the path to the results will 
always be clearly delineated. We refer the reader to the literature
reviewed in this Introduction for additional details; another
repository of relevant derivations is Martel's PhD dissertation
\cite{martel:phd}. Throughout this paper we adopt the sign conventions
of Misner, Thorne, and Wheeler \cite{MTW:73}, and we set $c=G=1$.    

\section{Schwarzschild spacetime}
\label{ii} 

The Schwarzschild metric is expressed as
\begin{equation}  
ds^2 = g_{ab}\, dx^a dx^b + r^2 \Omega_{AB}\, 
d\theta^A d\theta^B, 
\label{2.1}
\end{equation} 
in a form that is covariant under two-dimensional coordinate
transformations $x^a \to x^{\prime a}$. The coordinates $x^a$ span the 
submanifold ${\cal M}^2$ of the Schwarzschild spacetime --- the
``$(t,r)$ plane'' --- and lower-case Latin indices $a$, $b$, $c$,
etc.\ run over the values 0 and 1. The coordinates $\theta^A =
(\theta,\phi)$ span the two-spheres $x^a = \mbox{constant}$, and
upper-case Latin indices $A$, $B$, $C$, etc.\ run over the values 2
and 3. The full spacetime manifold is ${\cal M} = {\cal M}^2
\times S^2$. The two-dimensional tensor $g_{ab}$ and the scalar $r$
are functions of the coordinates $x^a$, and $\Omega_{AB} 
= \mbox{diag}(1,\sin^2\theta)$ is the metric on the unit two-sphere.  

We shall use three different coordinate systems $x^a$ in the
applications of the perturbation formalism to be presented below. The 
first is $(t,r)$, the usual Schwarzschild coordinates. The second is
$(u,r)$, where the retarded-time coordinate $u$ is defined by $u = 
t - r - 2M\ln(r/2M - 1)$. The third is $(v,r)$, where the
advanced-time coordinate $v$ is defined by $v = t + r 
+ 2M\ln(r/2M - 1)$. In these coordinates the Schwarzschild metric
takes the form   
\begin{eqnarray}
g_{ab}\, dx^a dx^b &=& f\, dt^2 + f^{-1} dr^2,  
\label{2.2} \\ 
&=& f\, du^2 - 2\, dudr,  
\label{2.3} \\ 
&=& f\, dv^2 + 2\, dvdr,
\label{2.4}
\end{eqnarray}
where $f := 1-2M/r$ and $M$ is the mass of the black hole. These
systems share the property that the scalar $r$ is adopted as one of
the coordinates. Our formalism is not, however, limited to these
coordinate choices; one retains the freedom of using any coordinate
system whatever, for example, harmonic coordinates, isotropic
coordinates, or double-null coordinates.    

We introduce the dual vector 
\begin{equation} 
r_a := \frac{\partial r}{\partial x^a}, 
\label{2.5}
\end{equation} 
which is normal to the surfaces of constant $r(x^a)$; in the
coordinates of Eqs.~(\ref{2.2})--(\ref{2.4}), $r_a = (0,1)$. We   
use $g^{ab}$, the inverse to $g_{ab}$, to raise its index: $r^a =
g^{ab} r_b$. This allows us to give a covariant definition to the
function $f$ that appears in Eqs.~(\ref{2.2})--(\ref{2.4}):   
\begin{equation}
r^a r_a =: f = 1 - \frac{2M}{r}. 
\label{2.6}
\end{equation} 
We also introduce $\varepsilon_{ab}$, the (antisymmetric) Levi-Civita 
tensor on ${\cal M}^2$; in the coordinates of
Eqs.~(\ref{2.2})--(\ref{2.4}), $\varepsilon_{tr} = \varepsilon_{ur} =
\varepsilon_{vr} = 1$. The timelike Killing vector of the
Schwarzschild spacetime is tangent to ${\cal M}^2$ and is given by  
\begin{equation}
t^a = -\varepsilon^{ab} r_b; 
\label{2.7}
\end{equation} 
in the coordinates of Eqs.~(\ref{2.2})--(\ref{2.4}), $t^a = (1,0)$. We 
have $t^a t_a = -f$ and $t^a r_a = 0$, and the vectors $r^a$, $t^a$
form a basis on ${\cal M}^2$. In terms of this basis we have $g^{ab} =
f^{-1}(-t^a t^b + r^a r^b)$ and $\varepsilon^{ab} = -f^{-1}(t^a r^b 
- r^a t^b)$.      

The covariant derivative operator compatible with $g_{ab}$ is denoted
$\nabla_a$; we thus have $\nabla_a g_{bc} \equiv 0$. It is easy to
show that for the Schwarzschild solution, 
\begin{equation} 
\nabla_a \nabla_b r = \frac{M}{r^2} g_{ab}, 
\label{2.8}
\end{equation}
so that $\Box r = 2M/r^2$, where $\Box := g^{ab} \nabla_a \nabla_b$ is  
the Laplacian operator on ${\cal M}^2$. We also have $\nabla_a t_b =
(M/r^2) \varepsilon_{ab}$, which confirms that $t^a$ is a Killing
vector. The Riemann tensor on ${\cal M}^2$ is $R_{abcd} =
(2M/r^3)(g_{ac} g_{bd} - g_{ad} g_{bc})$.  

We let $\Omega^{AB}$ be the inverse to $\Omega_{AB}$, the metric on
the unit two-sphere. The covariant derivative operator compatible with
$\Omega_{AB}$ is denoted $D_A$; we thus have $D_A \Omega_{BC} \equiv
0$. The Levi-Civita tensor on the unit two-sphere is denoted
$\varepsilon_{AB}$, and $\varepsilon_{\theta\phi} = \sin\theta$. The
Riemann tensor on the unit-sphere is $R_{ABCD} = \Omega_{AC}
\Omega_{BD} - \Omega_{AD} \Omega_{BD}$.  

Covariant differentiation in the Schwarzschild spacetime can be
defined in terms of covariant differentiation in the submanifolds
${\cal M}^2$ and $S^2$. If $\Gamma^a_{\ bc}$ is the connection
associated with $\nabla_a$, and if $\Gamma^A_{\ BC}$ is the connection
associated with $D_A$, then it is easy to show that the nonvanishing
components of the spacetime connection are given by 
$\mbox{}^4 \Gamma^a_{\ bc} = \Gamma^a_{\ bc}$,
$\mbox{}^4 \Gamma^a_{\ BC} = -r r^a \Omega_{BC}$, $\mbox{}^4
\Gamma^A_{\ Bc} = r^{-1} r_c \delta^A_{\ B}$, and $\mbox{}^4
\Gamma^A_{\ BC} = \Gamma^A_{\ BC}$. Using these rules we find that the
conservation identities for a stress-energy tensor $T^{\alpha\beta}$
in the Schwarzschild spacetime take the form 
\begin{equation} 
\nabla_b T^{ab} + D_B T^{aB} + \frac{2}{r} r_b T^{ab} 
- rr^a \Omega_{AB} T^{AB} = 0 
\label{2.9}
\end{equation}
and 
\begin{equation}
\nabla_a T^{aA} + D_B T^{AB} + \frac{4}{r} r_a T^{aA} = 0 
\label{2.10}
\end{equation}
when expressed in terms of the submanifold connections.    

\section{Spherical harmonics} 
\label{iii} 

In this section we introduce the scalar, vector, and tensor spherical 
harmonics that are used in the decomposition of the metric
perturbation. The tensorial nature of the spherical harmonics refers
to the unit two-sphere, and in this section we use the metric
$\Omega_{AB}$ and its inverse $\Omega^{AB}$ to lower and raise all 
upper-case Latin indices. All tensorial operations (including
covariant differentiation) shall refer to this metric. 

The scalar harmonics are the usual spherical-harmonic functions
$Y^{lm}(\theta^A)$. They satisfy the eigenvalue equation
$[\Omega^{AB} D_A D_B + l(l+1)] Y^{lm} = 0$.  

Vectorial spherical harmonics come in two types. The even-parity
harmonics are defined by 
\begin{equation} 
Y_{A}^{lm} := D_A Y^{lm}, 
\label{3.1}
\end{equation} 
while the odd-parity harmonics are
\begin{equation} 
X_{A}^{lm} := -\varepsilon_A^{\ B} D_B Y^{lm}. 
\label{3.2}
\end{equation} 
Their components are listed explicitly in Appendix \ref{A}. The 
vectorial harmonics satisfy the orthogonality relations   
\begin{equation} 
\int \bar{Y}^A_{lm} Y^{l'm'}_A\, d\Omega =  
l(l+1)\, \delta_{ll'}\delta_{mm'}
\label{3.3}
\end{equation}
and
\begin{equation} 
\int \bar{X}^A_{lm} X^{l'm'}_A\, d\Omega =  
l(l+1)\, \delta_{ll'}\delta_{mm'},
\label{3.4}
\end{equation}
in which an overbar indicates complex conjugation and $d\Omega :=
\sin\theta\, d\theta d\phi$ is an element of solid angle. We also have
\begin{equation} 
\int \bar{Y}^{A}_{lm} X^{l'm'}_A\, d\Omega = 0,
\label{3.5}
\end{equation} 
which states that the even-parity and odd-parity harmonics are always
orthogonal. The definitions (\ref{3.1}) and (\ref{3.2}) for the
vectorial spherical harmonics are identical to those provided by Regge 
and Wheeler \cite{regge-wheeler:57}. 

Tensorial spherical harmonics come in the same two types. The 
even-parity harmonics are $\Omega_{AB} Y^{lm}$ and 
\begin{equation}
Y_{AB}^{lm} := \Bigl[ D_A D_B + \frac{1}{2} l(l+1) \Omega_{AB}
\Bigr] Y^{lm}, 
\label{3.6}
\end{equation}
while the odd-parity harmonics are 
\begin{equation} 
X_{AB}^{lm} := -\frac{1}{2} \bigl( \varepsilon_A^{\ C} D_B +
\varepsilon_B^{\ C} D_A \bigr) D_C Y^{lm}. 
\label{3.7}
\end{equation} 
Their components are listed explicitly in Appendix \ref{A}. The
tensorial harmonics satisfy the orthogonality relations   
\begin{equation} 
\int \bar{Y}^{AB}_{lm} Y^{l'm'}_{AB}\, d\Omega =  
\frac{1}{2} (l-1)l(l+1)(l+2)\, \delta_{ll'}\delta_{mm'}
\label{3.8}
\end{equation}
and
\begin{equation} 
\int \bar{X}^{AB}_{lm} X^{l'm'}_{AB}\, d\Omega =  
\frac{1}{2} (l-1)l(l+1)(l+2)\, \delta_{ll'}\delta_{mm'}.
\label{3.9}
\end{equation}
We also have
\begin{equation} 
\int \bar{Y}^{AB}_{lm} X^{l'm'}_{AB}\, d\Omega = 0
\label{3.10}
\end{equation} 
and 
\begin{equation} 
\Omega^{AB} Y_{AB}^{lm} = 0 = \Omega^{AB} X_{AB}^{lm}. 
\label{3.11}
\end{equation} 
The definition (\ref{3.6}) for the even-parity harmonics does not
agree with that of Regge and Wheeler \cite{regge-wheeler:57}, who work
instead with the set $\Omega_{AB} Y^{lm}$ and $D_A D_B Y^{lm}$. We
find it more convenient to form the tracefree combinations
$Y_{AB}^{lm}$, which have the property of being (pointwise) orthogonal
to $\Omega_{AB} Y^{lm}$. The definition (\ref{3.7}) for the odd-parity 
harmonics also differs from Regge and Wheeler's, but only by an
overall minus sign, which we find convenient to introduce. 

The tensorial harmonics $Y_{AB}^{lm}$ and $X_{AB}^{lm}$ can be related 
to the spherical-harmonic functions of spin-weight $s=\pm 2$
\cite{goldberg-etal:67}, and to the pure-spin harmonics used by Thorne
\cite{thorne:80}. These relations are explored in Appendix \ref{A}.  

\section{Even-parity sector} 
\label{iv} 

\subsection{Perturbation fields and gauge transformations}  

The even-parity sector refers to those components of the metric
perturbation that can be expanded in terms of the even-parity
spherical harmonics $Y^{lm}$, $Y_A^{lm}$, $\Omega_{AB} Y^{lm}$, and
$Y_{AB}^{lm}$. Introducing the notation $\mbox{}^4 g_{ab} = g_{ab} +
p_{ab}$, $\mbox{}^4 g_{aB} = p_{aB}$, and $\mbox{}^4 g_{AB} = r^2
\Omega_{AB} + p_{AB}$ for the perturbed metric, the even-parity sector
of the metric perturbation is 
\begin{eqnarray} 
p_{ab} &=& \sum_{lm} h^{lm}_{ab} Y^{lm}, 
\label{4.1} \\ 
p_{aB} &=& \sum_{lm} j^{lm}_a Y_B^{lm},  
\label{4.2} \\ 
p_{AB} &=& r^2 \sum_{lm} \bigl( K^{lm} \Omega_{AB} Y^{lm} 
+ G^{lm} Y_{AB}^{lm} \bigr). 
\label{4.3}
\end{eqnarray} 
In this section the sums over $l$ are taken to begin at $l=2$; the low
multipoles ($l=0$ and $l=1$) will be considered separately in
Sec.~\ref{viii}. The fields $h^{lm}_{ab}$, $j^{lm}_{a}$, $K^{lm}$, and
$G^{lm}$ are defined on ${\cal M}^2$ and they depend on the coordinates
$x^a$ only. They are closely related to the quantities first
introduced by Regge and Wheeler \cite{regge-wheeler:57}, who worked
exclusively in terms of the usual Schwarzschild coordinates
$(t,r)$. In these coordinates (discarding for brevity the
spherical-harmonic labels) we have $h_{tt} = f H_0$, $h_{tr} = H_1$,
$h_{rr} = H_2/f$, $j_t = h_0$, and $j_r = h_1$. The function $G$
introduced in Eq.~(\ref{4.3}) is identical to the corresponding
Regge-Wheeler quantity, but $K$ is different: $K_{\rm here} 
= K_{\rm RW} - \frac{1}{2} l(l+1) G$; the difference originates from
the fact that Regge and Wheeler work with $\Omega_{AB} Y^{lm}$ and 
$D_A D_B Y^{lm}$ instead of $\Omega_{AB} Y^{lm}$ and $Y_{AB}^{lm}$.   

Even-parity gauge transformations are generated by a dual vector field
$\Xi_\alpha = (\Xi_a,\Xi_A)$ that is expanded as 
\begin{eqnarray} 
\Xi_a &=& \sum_{lm} \xi^{lm}_{a} Y^{lm}, 
\label{4.4} \\ 
\Xi_A &=& \sum_{lm} \xi^{lm} Y_A^{lm};
\label{4.5}
\end{eqnarray} 
the fields $\xi^{lm}_a$ and $\xi^{lm}$ depend on the coordinates $x^a$
only. Under such a transformation the perturbation quantities change
according to (see Appendix \ref{B}) 
\begin{eqnarray}
h_{ab} &\to& h'_{ab} = h_{ab} - \nabla_a \xi_b - \nabla_b \xi_a, 
\label{4.6} \\ 
j_{a} &\to& j'_{a} = j_a - \xi_a -\nabla_a \xi + \frac{2}{r} r_a \xi, 
\label{4.7} \\ 
K &\to& K' = K + \frac{l(l+1)}{r^2} \xi - \frac{2}{r} r^a \xi_a, 
\label{4.8} \\ 
G &\to& G' = G - \frac{2}{r^2} \xi,  
\label{4.9}
\end{eqnarray} 
where we have discarded the spherical-harmonic labels for brevity (we 
shall continue with this practice until the end of the section). It
is easy to show that the combinations 
\begin{equation} 
\tilde{h}_{ab} := h_{ab} - \nabla_a \varepsilon_b 
    - \nabla_b \varepsilon_a
\label{4.10}
\end{equation}
and 
\begin{equation} 
\tilde{K} := K + \frac{1}{2} l(l+1) G - \frac{2}{r} r^a \varepsilon_a  
\label{4.11}
\end{equation}
are gauge invariant, where 
\begin{equation}
\varepsilon_a := j_a - \frac{1}{2} r^2 \nabla_a G.  
\label{4.12}
\end{equation}
Equations (\ref{4.7}) and (\ref{4.9}) reveal that one can always 
choose a gauge in which $j_a = 0 = G$; this is the {\it Regge-Wheeler  
gauge}. Equations (\ref{4.10})--(\ref{4.12}) imply that
$\tilde{h}_{ab} = h_{ab}$ and $\tilde{K} = K$ in the Regge-Wheeler
gauge. 

\subsection{Perturbation equations} 

The Ricci tensor of the Schwarzschild spacetime vanishes, and as a
consequence its perturbation is gauge invariant. Its computation can
therefore be carried out in any convenient gauge, and the
Regge-Wheeler gauge is clearly convenient. The steps involved are as
follows. We substitute Eqs.~(\ref{4.1})--(\ref{4.3}), having set
$j_a = G = 0$, into the Ricci tensor of Appendix \ref{B}. We simplify  
the result and find that $\delta R_{ab}$ is expanded in terms of 
$Y^{lm}$, $\delta R_{aB}$ in terms of $Y_B^{lm}$, and $\delta R_{AB}$
in terms of both $\Omega_{AB} Y^{lm}$ and $Y_{AB}^{lm}$. From the
Ricci tensor we compute the Einstein tensor and set the result equal
to $8\pi T^{\alpha\beta}$, which describes the material source for the
gravitational perturbations. Each spherical-harmonic component of the
field equations can then be extracted by involving the orthonormality 
relations (\ref{3.3}) and (\ref{3.8}) satisfied by the spherical
harmonics. At the end of this calculation we take advantage of the
fact that $\tilde{h}_{ab} = h_{ab}$ and $\tilde{K} = K$ in the
Regge-Wheeler gauge. This allows us to make the substitutions $h_{ab}
\to \tilde{h}_{ab}$, $K \to \tilde{K}$ and therefore to express the
gauge-invariant Einstein tensor in terms of gauge-invariant
quantities.  

Our final results are      
\begin{widetext} 
\begin{eqnarray} 
Q_{ab} &=& \nabla_c \nabla_{(a} \tilde{h}^c_{\ b)} 
- \frac{1}{2} g_{ab} \nabla_c \nabla_d \tilde{h}^{cd} 
- \frac{1}{2} \Box \bigl( \tilde{h}_{ab} - g_{ab} \tilde{h} \bigr) 
- \frac{1}{2} \nabla_a \nabla_b \tilde{h} 
+ \frac{2}{r} r_c \bigl( \nabla_{(a} \tilde{h}^c_{\ b)} 
                    - g_{ab} \nabla_d \tilde{h}^{cd} \bigr) 
\nonumber \\ & & \mbox{}
- \frac{1}{r} r^c \nabla_c \bigl( \tilde{h}_{ab} 
- g_{ab} \tilde{h} \bigr)  
+ \frac{l(l+1)}{2 r^2} \tilde{h}_{ab} 
- \frac{1}{r^2} g_{ab} r_c r_d \tilde{h}^{cd} 
- \frac{1}{2} \biggl[ \frac{l(l+1)}{r^2} + \frac{2M}{r^3} \biggr] 
    g_{ab} \tilde{h}  
\nonumber \\ & & \mbox{}
- \nabla_a \nabla_b \tilde{K} + g_{ab} \Box \tilde{K} 
- \frac{2}{r} r_{(a} \nabla_{b)} \tilde{K} 
+ \frac{3}{r} g_{ab} r^c \nabla_c \tilde{K} 
- \frac{(l-1)(l+2)}{2r^2} g_{ab} \tilde{K}, 
\label{4.13} \\ 
Q_{a} &=& \nabla_c \tilde{h}^c_{\ a} - \nabla_a \tilde{h} 
+ \frac{1}{r} r_a \tilde{h} - \nabla_a \tilde{K}, 
\label{4.14} \\ 
Q^\flat &=& \Box \tilde{h} - \nabla_a \nabla_b \tilde{h}^{ab} 
- \frac{2}{r} r_a \nabla_b \tilde{h}^{ab} 
+ \frac{1}{r} r^a \nabla_a \tilde{h} 
- \frac{l(l+1)}{2 r^2} \tilde{h} 
+ \Box \tilde{K} + \frac{2}{r} r^a \nabla_a \tilde{K}, 
\label{4.15} \\
Q^\sharp &=& -\tilde{h}, 
\label{4.16}
\end{eqnarray}
\end{widetext} 
where $\tilde{h} := g^{ab} \tilde{h}_{ab}$ and $\Box := g^{ab}
\nabla_a \nabla_b$. The source terms are 
\begin{eqnarray}  
Q^{ab} &=& 8\pi \int T^{ab} \bar{Y}^{lm}\, d\Omega, 
\label{4.17} \\
Q^a &=& \frac{16\pi r^2}{l(l+1)} \int T^{aB} \bar{Y}^{lm}_B\, d\Omega, 
\label{4.18} \\ 
Q^\flat &=& 8\pi r^2 \int T^{AB} \Omega_{AB} \bar{Y}^{lm}\, d\Omega, 
\label{4.19} \\ 
Q^\sharp &=& \frac{32\pi r^4}{(l-1)l(l+1)(l+2)} \int T^{AB}
\bar{Y}^{lm}_{AB}\, d\Omega. \qquad 
\label{4.20}
\end{eqnarray} 
In Eqs.~(\ref{4.17})--(\ref{4.20}) the stress-energy tensor is
imagined to be given in fully contravariant form; $T^{ab}$,
$T^{aB}$, and $T^{AB}$ are then its components in the spacetime
coordinates $(x^a,\theta^A)$. In the event where the stress-energy
tensor would be given in covariant or mixed form, its indices would
have to be raised with $(g^{ab},r^{-2} \Omega^{AB})$ --- the inverse
Schwarzschild metric --- before evaluating the source terms. In 
Eqs.~(\ref{4.13})--(\ref{4.16}) all lower-case Latin indices are
lowered and raised with $g_{ab}$ and $g^{ab}$, respectively.    

The perturbation equations are not all independent. By virtue of the
Bianchi identities, or the conservation equations (\ref{2.9}) and 
(\ref{2.10}), they are related by  
\begin{equation} 
\nabla_b Q^{ab} + \frac{2}{r} r_b Q^{ab} - \frac{l(l+1)}{2r^2} Q^a
- \frac{1}{r} r^a Q^\flat = 0
\label{4.21}
\end{equation}
and 
\begin{equation} 
\nabla_a Q^a + \frac{2}{r} r_a Q^a - \frac{(l-1)(l+2)}{2 r^2} Q^\sharp
+ Q^\flat = 0.  
\label{4.22}
\end{equation}  

\subsection{Master equation} 

The {\it Zerilli-Moncrief function} is defined by 
\begin{equation}
\Psi^{lm}_{\rm even} := \frac{2r}{l(l+1)} \biggl[ \tilde{K}^{lm}  
+ \frac{2}{\Lambda} \bigl( r^a r^b \tilde{h}^{lm}_{ab} 
- r r^a \nabla_a \tilde{K}^{lm} \bigr) \biggr],  
\label{4.23}
\end{equation}
where  
\begin{equation}
\Lambda := (l-1)(l+2) + \frac{6M}{r}. 
\label{4.24}
\end{equation} 
This is a covariant and gauge-invariant generalization of
the definition provided by Lousto and Price \cite{lousto-price:97},
who worked with the usual Schwarzschild coordinates and in the
Regge-Wheeler gauge. Their even-parity function agrees, up to a
normalization factor, with the gauge-invariant function first
introduced by Moncrief \cite{moncrief:74}. Moncrief's function, in
turn, is a variant of Zerilli's original even-parity function
\cite{zerilli:70} (Zerilli worked in the frequency domain instead of
the Schwarzschild time domain). Up to the same normalization factor
the Moncrief and Fourier-transformed Zerilli functions are equal up to
the presence of source terms; they are equal only where there is no
matter. The normalization adopted in Eq.~(\ref{4.23}) will be seen to
be convenient when we discuss gravitational radiation at future null
infinity (Sec.~\ref{vi}) and at the horizon (Sec.~\ref{vii}); our
definition of the even-parity master function is well adapted to the
description of radiation fields. 

The perturbation equations (\ref{4.13})--(\ref{4.16}) give rise to a
wave equation for the function $\Psi_{\rm even}$ (we resume our
practice of omitting the spherical-harmonic labels). The manipulations
are long and tedious, and we shall not present them here (the reader
may consult Ref.~\cite{martel:phd} for details). We simply state the
final result: As a consequence of the field equations, the even-parity
master function satisfies the {\it Zerilli equation} 
\begin{equation} 
( \Box - V_{\rm even} ) \Psi_{\rm even} = S_{\rm even}, 
\label{4.25}
\end{equation} 
with potential 
\begin{equation} 
V_{\rm even} = \frac{1}{\Lambda^2} \biggl[ \mu^2 \biggl(
  \frac{\mu+2}{r^2} + \frac{6M}{r^3} \biggr) 
+ \frac{36M^2}{r^4} \biggl(\mu + \frac{2M}{r} \biggr) \biggr] 
\label{4.26}
\end{equation}
and source term 
\begin{widetext} 
\begin{equation}
S_{\rm even} = \frac{4}{\Lambda} r_a Q^a - \frac{1}{r} Q^\sharp 
+ \frac{2}{(\mu+2)\Lambda} \biggl\{ - 2r^2 r^a \nabla_a Q +
\frac{24M}{\Lambda} r_a r_b Q^{ab} + 2r f Q^\flat + \frac{r}{\Lambda} 
\biggl[ \mu(\mu-2) + 12(\mu-3) \frac{M}{r} + 84\frac{M^2}{r^2} \biggr] 
Q \biggr\}, 
\label{4.27}
\end{equation}
\end{widetext} 
where $\mu := (l-1)(l+2)$ and $Q := g_{ab} Q^{ab}$. The validity of 
Eqs.~(\ref{4.25})--(\ref{4.27}) can be verified by brute-force
evaluation of both sides of Eqs.~(\ref{4.25}). The general source 
term for the Zerilli equation has never been presented in the
literature; we display it here for the first time, in a form covariant
under an arbitrary transformation of the coordinates $x^a$ used in the 
two-dimensional submanifold ${\cal M}^2$ of the Schwarzschild spacetime. 
    
\section{Odd-parity sector} 
\label{v} 

\subsection{Perturbation fields and gauge transformations}  

The odd-parity sector refers to those components of the metric
perturbation that can be expanded in terms of the odd-parity
spherical harmonics $X_A^{lm}$ and $X_{AB}^{lm}$. Recalling the 
notation $\mbox{}^4 g_{ab} = g_{ab} + p_{ab}$, $\mbox{}^4 g_{aB} 
= p_{aB}$, and $\mbox{}^4 g_{AB} = r^2 \Omega_{AB} + p_{AB}$ for the
perturbed metric, the odd-parity sector of the metric perturbation is  
\begin{eqnarray} 
p_{ab} &=& 0, 
\label{5.1} \\ 
p_{aB} &=& \sum_{lm} h^{lm}_a X_B^{lm},  
\label{5.2} \\ 
p_{AB} &=& \sum_{lm} h^{lm}_2 X_{AB}^{lm}. 
\label{5.3}
\end{eqnarray} 
In this section the sums over $l$ are taken to begin at $l=2$; there
is no odd-parity perturbation with $l=0$, and the case $l=1$ will be
considered separately in Sec.~\ref{viii}. The fields $h^{lm}_{a}$ and
$h_2^{lm}$ are defined on ${\cal M}^2$ and they depend on the
coordinates $x^a$ only. They are closely related to the quantities
first introduced by Regge and Wheeler \cite{regge-wheeler:57}, who
worked exclusively in terms of the usual Schwarzschild coordinates
$(t,r)$. In these coordinates (discarding spherical-harmonic labels)
we have $h_{t} = h_0$ and $h_{r} = h_1$. Except for a sign --- refer
back to the discussion following Eq.~(\ref{3.11}) --- the function
$h_2$ is identical to the corresponding Regge-Wheeler quantity. 

Odd-parity gauge transformations are generated by a dual vector field 
$\Xi_\alpha = (0,\Xi_A)$ that is expanded as 
\begin{equation} 
\Xi_A = \sum_{lm} \xi^{lm} X_A^{lm}, 
\label{5.4}
\end{equation} 
in which $\xi^{lm}$ depends on the coordinates $x^a$ only. Under such
a transformation the perturbation quantities change according to (see
Appendix \ref{B}) 
\begin{eqnarray}
h_{a} &\to& h'_{a} = h_{a} - \nabla_a \xi + \frac{2}{r} r_a \xi,  
\label{5.5} \\ 
h_2 &\to& h'_2 = h_2 - 2\xi, 
\label{5.6}
\end{eqnarray} 
where we have discarded the spherical-harmonic labels for brevity (we 
shall continue with this practice until the end of the section). It
is easy to show that the combinations 
\begin{equation} 
\tilde{h}_a = h_a - \frac{1}{2} \nabla_a h_2 + \frac{1}{r} r_a h_2  
\label{5.7} 
\end{equation}
are gauge invariant. Equation (\ref{5.6}) reveals that one can always  
choose a gauge in which $h_2 = 0$; this is the {\it Regge-Wheeler
gauge}. Equation (\ref{5.7}) implies that $\tilde{h}_{a} = h_{a}$
in the Regge-Wheeler gauge.  

\subsection{Perturbation equations} 

The Ricci tensor of the Schwarzschild spacetime vanishes, and as a
consequence its perturbation is gauge invariant. Its computation can
therefore be carried out in any convenient gauge, and as in the
preceding section we shall adopt the Regge-Wheeler gauge. We
substitute Eqs.~(\ref{5.1})--(\ref{5.3}), having set $h_2 = 0$, into
the Ricci tensor of Appendix \ref{B}. We simplify the result and find
that $\delta R_{ab}$ vanishes, $\delta R_{aB}$ is expanded in terms of
$X_B^{lm}$, and $\delta R_{AB}$ in terms of $X_{AB}^{lm}$. From the
Ricci tensor we compute the Einstein tensor and set the result equal
to $8\pi T^{\alpha\beta}$. Each spherical-harmonic component of the
field equations can then be extracted by involving the orthonormality 
relations (\ref{3.4}) and (\ref{3.9}) satisfied by the odd-parity
harmonics. At the end of this calculation we take advantage of the
fact that $\tilde{h}_{a} = h_{a}$ in the Regge-Wheeler gauge. This
allows us to make the substitution $h_{a} \to \tilde{h}_{a}$ and
therefore to express the gauge-invariant Einstein tensor in terms of
gauge-invariant quantities.  

Our final results are      
\begin{eqnarray} 
P_a &=& -\Box \tilde{h}_a + \nabla_a \nabla_b \tilde{h}^b 
+ \frac{2}{r} \bigl( r_b \nabla_a \tilde{h}^b 
                   - r_a \nabla_b \tilde{h}^b \bigr)
\nonumber \\ & & \mbox{}
- \frac{2}{r^2} r_a r_b \tilde{h}^b 
+ \frac{l(l+1)}{r^2} \tilde{h}_a, 
\label{5.8} \\ 
P &=& \nabla_a \tilde{h}^a, 
\label{5.9}
\end{eqnarray} 
where $\Box := g^{ab} \nabla_a \nabla_b$. The source terms are 
\begin{eqnarray}  
P^a &=& \frac{16\pi r^2}{l(l+1)} \int T^{aB} \bar{X}^{lm}_B\, d\Omega,  
\label{5.10} \\ 
P &=& \frac{16\pi r^4}{(l-1)l(l+1)(l+2)} \int T^{AB}
\bar{X}^{lm}_{AB}\, d\Omega. \qquad 
\label{5.11}
\end{eqnarray} 
In Eqs.~(\ref{5.10}) and (\ref{5.11}) the stress-energy tensor is 
imagined to be given in fully contravariant form; $T^{aB}$ and  
$T^{AB}$ are its relevant components in the spacetime coordinates 
$(x^a,\theta^A)$. In the event where the stress-energy tensor would be 
given in covariant or mixed form, its indices would have to be raised
with $(g^{ab},r^{-2} \Omega^{AB})$ --- the inverse Schwarzschild
metric --- before evaluating the source terms. In 
Eqs.~(\ref{5.8}) and (\ref{5.9}) all lower-case Latin indices are 
lowered and raised with $g_{ab}$ and $g^{ab}$, respectively.    

The perturbation equations are not all independent. By virtue of the
Bianchi identities, or the conservation equation (\ref{2.10}), they
are related by   
\begin{equation} 
\nabla_a P^a + \frac{2}{r} r_a P^a - \frac{(l-1)(l+2)}{r^2} P = 0. 
\label{5.12}
\end{equation} 

\subsection{Master equation} 

The {\it Cunningham-Price-Moncrief function} is defined by 
\begin{equation}
\Psi^{lm}_{\rm odd} := \frac{2r}{(l-1)(l+2)} \varepsilon^{ab} 
\Bigl( \nabla_a \tilde{h}^{lm}_b - \frac{2}{r} r_a \tilde{h}^{lm}_b
\Bigr),  
\label{5.13}
\end{equation}
where $\varepsilon_{ab}$ is the Levi-Civita tensor on the submanifold 
${\cal M}^2$. Apart from a different normalization factor, this is a
covariant generalization of the definition provided by Cunningham,
Price, and Moncrief \cite{cunningham-etal:78}, who worked with the
usual Schwarzschild coordinates. Our definition (and normalization)
agrees with the odd-parity function considered by Jhingan and Tanaka 
\cite{jhingan-tanaka:03}, who also worked with the Schwarzschild
coordinates, but in the frequency domain. The normalization adopted in
Eq.~(\ref{5.13}) will be seen to be convenient when we discuss
gravitational radiation at future null infinity (Sec.~\ref{vi}) and at
the horizon (Sec.~\ref{vii}); our definition of the odd-parity master
function is well adapted to the description of radiation fields. It is
noteworthy that it can also be expressed as   
\[
\Psi^{lm}_{\rm odd} := \frac{2r}{(l-1)(l+2)} \varepsilon^{ab} 
\Bigl( \partial_a h^{lm}_b - \frac{2}{r} r_a h^{lm}_b \Bigr), 
\]
in terms of the original (gauge-dependent) perturbation quantities, 
and in terms of the partial differentiation operator; the function is
nonetheless gauge-invariant and a scalar.   

The perturbation equations (\ref{5.8}) and (\ref{5.9}) give rise to 
a wave equation for the function $\Psi_{\rm odd}$ (we resume our
practice of discarding the spherical-harmonic labels). As in the
preceding section we simply state the final result: As a consequence
of the field equations, the odd-parity master function satisfies the 
{\it Regge-Wheeler equation}  
\begin{equation} 
( \Box - V_{\rm odd} ) \Psi_{\rm odd} = S_{\rm odd}, 
\label{5.14}
\end{equation} 
with potential 
\begin{equation} 
V_{\rm odd} = \frac{l(l+1)}{r^2} - \frac{6M}{r^3} 
\label{5.15}
\end{equation}
and source term 
\begin{equation}
S_{\rm odd} = - \frac{2r}{(l-1)(l+2)} \varepsilon^{a b} \nabla_a P_b. 
\label{5.16}
\end{equation} 
The validity of Eqs.~(\ref{5.14})--(\ref{5.16}) can be verified by
brute-force evaluation of both sides of Eqs.~(\ref{5.14}). The general
source term for the Regge-Wheeler equation has never been presented in 
covariant form in the literature. We display it here for the first
time, and note that in the usual Schwarzschild coordinates (and in the
frequency domain), $S_{\rm odd}$ agrees with the source term presented
in Eq.~(18) of Jhingan and Tanaka \cite{jhingan-tanaka:03}.  

\subsection{Regge-Wheeler function} 

The Cunningham-Price-Moncrief function is a close cousin to the more
familiar {\it Regge-Wheeler function} \cite{regge-wheeler:57}, whose
covariant and gauge-invariant definition is 
\begin{equation}
\Psi^{lm}_{\rm RW} := \frac{1}{r} r^a \tilde{h}^{lm}_a.  
\label{5.17}
\end{equation}
As we shall see in Secs.~\ref{vi} and \ref{vii}, the Regge-Wheeler
function is not well suited to describe the gravitational radiation 
field, and in this paper we adopt the function $\Psi^{lm}_{\rm odd}$
of Eq.~(\ref{5.13}) as the fundamental odd-parity master function. It
is straightforward to use the perturbation equations to show that
these functions are related by 
\begin{equation} 
\Psi_{\rm RW} = \frac{1}{2} t^a \nabla_a \Psi_{\rm odd} 
+ \frac{r}{(l-1)(l+2)} r_a P^a, 
\label{5.18}
\end{equation}
where $t^a = -\varepsilon^{ab} r_b$ is the Killing vector of
Eq.~(\ref{2.7}). Outside of sources, and apart from a factor of
one-half, the Regge-Wheeler function is the time derivative of the 
Cunningham-Price-Moncrief function.   

The function of Eq.~(\ref{5.17}) also satisfies the Regge-Wheeler
equation, 
\begin{equation} 
( \Box - V_{\rm odd} ) \Psi_{\rm RW} = S_{\rm RW}, 
\label{5.19}
\end{equation} 
with the potential of Eq.~(\ref{5.15}) and a new source term given by 
\begin{equation}
S_{\rm RW} = \frac{1}{r} \biggl[ r^a \bigl( \nabla_a P - P_a \bigr) 
- \frac{2}{r} \biggl( 1 - \frac{3M}{r} \biggr) P \biggr]. 
\label{5.20} 
\end{equation} 
This covariant form for the source term is also a new result.  

\section{Radiation at future null infinity} 
\label{vi} 

To examine the gravitational perturbations near future null infinity
we adopt the retarded coordinates $(u,r,\theta,\phi)$ and express the 
two-dimensional Schwarzschild metric in the form of
Eq.~(\ref{2.3}). In these coordinates, future null infinity
corresponds to taking the limit $r \to \infty$ keeping $u$ fixed, and
our strategy will be to expand the metric perturbations in powers of 
$r^{-1}$. In asymptotically Cartesian coordinates the radiative part
of the metric would scale as $r^{-1}$; transforming to spherical
coordinates produces the scalings
\begin{equation}
p^{\rm rad}_{ab} = O(r^{-1}), \quad
p^{\rm rad}_{aB} = O(r^0), \quad
p^{\rm rad}_{AB} = O(r)
\label{6.1}
\end{equation} 
for the radiative part of the metric perturbations. Our goal is 
to isolate this, and to calculate how much energy and angular momentum
is carried away by the radiation. We find it advantageous to work in
gauge in which  
\begin{equation}
t^a p_{ab} = 0 = t^a p_{a B}, 
\label{6.2}
\end{equation}
where $t^a$ is the Killing vector of Eq.~(\ref{2.7}). In spite of the
fact that $t^a$ is not a null vector (except on the even horizon,
which is well outside our domain of consideration), this gauge happens
to be a perfectly respectable ``radiation gauge.''      

In this section we take $l \geq 2$; as is well known, the low
multipoles $l=0$ and $l=1$ do not contain radiative degrees of
freedom. We assume that the matter distribution responsible for the
radiation is confined to a bounded volume, and that our domain of
consideration is outside this volume; we shall therefore be solving
the {\it vacuum} field equations. 

We begin with the even-parity sector of Sec.~\ref{iv}. The gauge
conditions imply $h_{uu} = h_{ur} = j_u = 0$, and the scalings of
Eq.~(\ref{6.1}) imply that we are looking for the $r^{-1}$ part of
$h_{rr}$, $K$, and $G$, as well as the $r^0$ part of $j_r$. These can
be determined by expanding the components of the metric perturbation
in powers of $r^{-1}$ and substituting them into the field equations
of Eqs.~(\ref{4.13})--(\ref{4.16}). The expansions for $h_{rr}$, $K$,
and $G$ begin at order $r^{-1}$ and each coefficient is a
to-be-determined function of $u$; the expansion for $j_r$ begins
instead at order $r^0$. The field equations return $Q_{ab}$, $Q_{a}$,
$Q^\flat$, and $Q^\sharp$ expanded in powers of $r^{-1}$, and setting
each coefficient to zero determines the metric perturbation. We obtain  
\begin{eqnarray} 
h_{rr} &=& -\frac{l(l+1)}{r^3} \int^u a(u')\, du' + O(r^{-4}),  
\label{6.3} \\  
j_r &=& \frac{a(u)}{r} + O(r^{-2}), 
\label{6.4} \\ 
K &=& \frac{l(l+1)}{2r^3} \int^u a(u')\, du' + O(r^{-4}), 
\label{6.5} \\ 
G &=& -\frac{2}{(l-1)(l+2)} \frac{\dot{a}(u)}{r} + O(r^{-2}), 
\label{6.6}
\end{eqnarray} 
where $a(u)$ is a function that is not determined by the vacuum field
equations, and $\dot{a} := da/du$. We see that the radiative part of
the perturbation is contained entirely in the function $G(u,r)$. With
the metric perturbation of Eqs.~(\ref{6.3})--(\ref{6.6}) we may
evaluate the Zerilli-Moncrief function of Eq.~(\ref{4.23}). The result
is  
\begin{equation} 
\Psi_{\rm even} = -\frac{2}{(l-1)(l+2)} \dot{a}(u) + O(r^{-1}), 
\label{6.7}
\end{equation}
and we conclude that the radiative part of the even-parity sector is
given by 
\begin{equation} 
p^{\rm rad,\ even}_{AB} = r \sum_{lm} \Psi^{lm}_{\rm even}(u,r=\infty)  
Y^{lm}_{AB}. 
\label{6.8}
\end{equation} 
It is obtained by integrating the Zerilli equation (\ref{4.25}) and
evaluating the Zerilli-Moncrief function at $r = \infty$.    

We turn next to the odd-parity sector of Sec.~\ref{v}. The gauge
conditions imply $h_{u} = 0$, and the scalings of Eq.~(\ref{6.1})
imply that we are looking for the $r^{0}$ part of $h_r$ and the 
$r^1$ part of $h_2$. These are determined by following the same
procedure as in the even-parity case, and we obtain 
\begin{eqnarray} 
h_{r} &=& -\frac{(l-1)(l+2)}{2 r} \int^u b(u')\, du' + O(r^{-2}),   
\label{6.9} \\  
h_2 &=& b(u) r + O(r^0), 
\label{6.10} 
\end{eqnarray} 
where $b(u)$ is not determined by the vacuum field equations. We see
that the radiative part of the perturbation is contained entirely in
the function $h_2(u,r)$. With the metric perturbation of 
Eqs.~(\ref{6.9}), (\ref{6.10}) we may evaluate the
Cunningham-Price-Moncrief function of Eq.~(\ref{5.13}). The result is   
\begin{equation} 
\Psi_{\rm odd} = b(u) + O(r^{-1}),  
\label{6.11}
\end{equation}
and we conclude that the radiative part of the odd-parity sector is
given by 
\begin{equation} 
p^{\rm rad,\ odd}_{AB} = r \sum_{lm} \Psi^{lm}_{\rm odd}(u,r=\infty) 
X^{lm}_{AB}. 
\label{6.12}
\end{equation} 
It is obtained by integrating the Regge-Wheeler equation (\ref{5.14})
and evaluating the Cunningham-Price-Moncrief function at $r =
\infty$. Notice that by virtue of Eq.~(\ref{5.18}), the radiative
field could instead be expressed in terms of the $u$-integral of the
Regge-Wheeler function. The need to perform this integration is
inconvenient, and it is the simple relationship of Eq.~(\ref{6.12})
that has motivated the adoption of the Cunningham-Price-Moncrief
function as the fundamental odd-parity master function.  

The full radiative field is obtained from Eqs.~(\ref{6.8}) and
(\ref{6.12}). We have $p^{\rm rad}_{ab} = 0 = p^{\rm rad}_{aB}$ and 
\begin{equation} 
p^{\rm rad}_{AB} = r \sum_{lm} \Bigl( \Psi^{lm}_{\rm even}    
Y^{lm}_{AB} + \Psi^{lm}_{\rm odd} X^{lm}_{AB} \Bigr), 
\label{6.13}
\end{equation} 
where $\Psi^{lm}_{\rm even} \equiv \Psi^{lm}_{\rm even}(u,r=\infty)$ 
and $\Psi^{lm}_{\rm odd} \equiv \Psi^{lm}_{\rm odd}(u,r=\infty)$. 
As expected, the radiative field is transverse, and tracefree by
virtue of Eqs.~(\ref{3.11}). The two fundamental polarizations of the 
gravitational wave can be defined by $h_+ := p_{\theta\theta}/r^2$ 
and $h_\times := p_{\theta\phi}/(r^2\sin\theta)$. Using the components
of the tensorial spherical harmonics listed in Appendix \ref{A}, we
obtain 
\begin{eqnarray}
h_+ &=& \frac{1}{r} \sum_{lm} \biggl\{ \Psi^{lm}_{\rm even}   
\biggl[ \frac{\partial^2}{\partial \theta^2} 
+ \frac{1}{2} l(l+1) \biggr] Y^{lm} 
\nonumber \\ & & \mbox{} 
- \Psi^{lm}_{\rm odd} 
\frac{i m}{\sin\theta} \biggl[ \frac{\partial}{\partial \theta} -
\frac{\cos\theta}{\sin\theta} \biggr] Y^{lm} \biggr\} 
\label{6.14}
\end{eqnarray} 
and 
\begin{eqnarray}
h_\times &=& \frac{1}{r} \sum_{lm} \biggl\{ \Psi^{lm}_{\rm even}   
\frac{i m}{\sin\theta} \biggl[ \frac{\partial}{\partial \theta} -
\frac{\cos\theta}{\sin\theta} \biggr] Y^{lm}
\nonumber \\ & & \mbox{} 
+ \Psi^{lm}_{\rm odd} \biggl[ \frac{\partial^2}{\partial \theta^2}  
+ \frac{1}{2} l(l+1) \biggr] Y^{lm} \biggr\} 
\label{6.15}. 
\end{eqnarray} 
The energy and angular momentum carried away by the gravitational
radiation can be calculated using the techniques developed by
Thorne \cite{thorne:80}. We compare our Eq.~(\ref{6.13}) to his
Eq.~(4.3), taking into account the relationship between our tensorial
harmonics and his pure-spin harmonics (this is spelled out in Appendix 
\ref{A}). Thorne's mass multipole moments are thus seen to be
proportional to $\Psi^{lm}_{\rm even}(u,r=\infty)$, and his current
moments are proportional to 
$\Psi^{lm}_{\rm odd}(u,r=\infty)$. Substituting these into Thorne's
Eq.~(4.16) we obtain 
\begin{eqnarray}
\biggl\langle \frac{dE}{du} \biggr\rangle &=& \frac{1}{64\pi} 
\sum_{lm} (l-1)l(l+1)(l+2)  
\nonumber \\ & & \mbox{} \times 
\Bigl\langle \bigl| \dot{\Psi}^{lm}_{\rm even} \bigr|^2
+ \bigl| \dot{\Psi}^{lm}_{\rm odd} \bigr|^2 \Bigr\rangle 
\label{6.16}
\end{eqnarray}
for the averaged rate at which the energy escapes to future null
infinity. Substituting instead into Thorne's Eq.~(4.23) returns 
\begin{eqnarray}
\biggl\langle \frac{dJ}{du} \biggr\rangle &=& \frac{1}{64\pi} 
\sum_{lm} (l-1)l(l+1)(l+2) (im) 
\nonumber \\ & & \mbox{} \times 
\Bigl\langle \bar{\Psi}^{lm}_{\rm even} \dot{\Psi}^{lm}_{\rm even}  
+ \bar{\Psi}^{lm}_{\rm odd} \dot{\Psi}^{lm}_{\rm odd} \Bigr\rangle  
\label{6.17}
\end{eqnarray}
for the averaged rate at which the angular momentum escapes to
infinity. This is the component of the angular-momentum vector in the
arbitrary $z$ direction which defines the orientation of the angles
$\theta$ and $\phi$. The overbar indicates complex conjugation, and it
is not difficult to show that $\langle dJ/du \rangle$ is real. The
averaging carried out in Eqs.~(\ref{6.16}) and (\ref{6.17}) is over a
characteristic time scale associated with the gravitational wave. 

\section{Radiation at the event horizon} 
\label{vii} 

To examine the gravitational perturbations near the event horizon we
adopt the advanced coordinates $(v,r,\theta,\phi)$ and express the
two-dimensional Schwarzschild metric in the form of
Eq.~(\ref{2.4}). We want to calculate how much energy and angular 
momentum is transfered to the black hole by the perturbation, and we 
shall do so by following the methods devised by Poisson
\cite{poisson:04c}. As in the preceding section we impose the gauge
conditions   
\begin{equation}
t^a p_{ab} = 0 = t^a p_{a B} 
\label{7.1}
\end{equation}
on the metric perturbation; $t^a$ is still the Killing vector of  
Eq.~(\ref{2.7}). Recall that this vector is null on the event horizon,  
and Eq.~(\ref{7.1}), evaluated at $r = 2M$, is equivalent to Poisson's
Eq.~(6.5). Poisson then shows that the part of the metric perturbation 
which is associated with the transport of energy and angular momentum
across the horizon is $p_{AB}$. [Refer to Poisson's Eq.~(6.10), which
establishes the equality between $p_{AB}$ and $\gamma^1_{AB}$, the
perturbation of the horizon's intrinsic metric.] 

In this section we take $l \geq 2$, as the low multipoles $l=0$ and
$l=1$ do not contain radiative degrees of freedom. We assume that the 
matter distribution responsible for the radiation does not come near 
the event horizon; we shall therefore be solving the {\it vacuum}
field equations in an empty neighborhood of the event horizon.   

We begin with the even-parity sector of Sec.~\ref{iv}. The gauge
conditions of Eq.~(\ref{7.1}) imply that $h_{vv} = h_{vr} = j_v = 0$,
so that $h_{rr}$, $j_r$, $K$, and $G$ are the only nonvanishing
components of the metric perturbation. As stated above, the radiation
crossing the event horizon is described entirely by $K$ and $G$
evaluated at $r=2M$. The field equations, however, imply that $K$
vanishes on the horizon. This is verified by expanding $h_{rr}$,
$j_r$, $K$, and $G$ in powers of $f = 1-2M/r$ and substituting them
into the field equations of Eqs.~(\ref{4.13})--(\ref{4.16}). This
calculation reveals also that $G(v,r=2M) = a(v)$, a function that is
not determined by the vacuum field equations. On the other hand, the
expansions allow us to evaluate the Zerilli-Moncrief function of
Eq.~(\ref{4.23}), and the result is $\Psi_{\rm even}(r=2M) = 2M
a(v)$. We conclude that on the horizon, the radiative part of the
even-parity sector is given by    
\begin{equation} 
p^{\rm rad,\ even}_{AB} = 2M \sum_{lm} \Psi^{lm}_{\rm even}(v,r=2M)  
Y^{lm}_{AB}. 
\label{7.2}
\end{equation} 
It is obtained by integrating the Zerilli equation (\ref{4.25}) and
evaluating the Zerilli-Moncrief function at $r = 2M$.    

We turn next to the odd-parity sector of Sec.~\ref{v}. The gauge
conditions imply $h_{u} = 0$, so that $h_r$ and $h_2$ are the only
nonvanishing components of the metric perturbation. The radiation
crossing the event horizon is described entirely by $h_2$ evaluated at
$r=2M$, and the field equations imply that $b(v) := h_2(v,r=2M)$
remains as an undetermined function. They also imply that $h_r(v,r=2M)
= c(v)$, with   
\[
c(v) := \frac{(l-1)(l+2)}{8 M^2} \int^v b(v')\, dv'.
\]
These statements are verified by expanding $h_r$ and $h_2$ in powers
of $f = 1-2M/r$ and substituting them into Eqs.~(\ref{5.8}) and
(\ref{5.9}). The expansions allow us also to evaluate the
Cunningham-Price-Moncrief function of Eq.~(\ref{5.13}), 
and the result is  
\[
\Psi_{\rm odd}(r=2M) = -\frac{4M}{(l-1)(l+2)} \frac{dc}{dv} 
= -\frac{1}{2M} b(v). 
\]
We conclude that on the horizon, the radiative part of the odd-parity
sector is given by
\begin{equation} 
p^{\rm rad,\ odd}_{AB} = -2M \sum_{lm} \Psi^{lm}_{\rm odd}(v,r=2M)  
X^{lm}_{AB}. 
\label{7.3}
\end{equation} 
It is obtained by integrating the Regge-Wheeler equation (\ref{5.14})
and evaluating the Cunningham-Price-Moncrief function at $r =
2M$. Notice that by virtue of Eq.~(\ref{5.18}), the radiative field
could instead be expressed in terms of the $v$-integral of the
Regge-Wheeler function; this would give rise to Poisson's Eq.~(7.3), a
less convenient expression.  

The full radiative field is obtained by adding Eqs.~(\ref{7.2}) and
(\ref{7.3}). We have  
\begin{equation} 
p^{\rm rad}_{AB} = 2M \sum_{lm} \Bigl( \Psi^{lm}_{\rm even}    
Y^{lm}_{AB} - \Psi^{lm}_{\rm odd} X^{lm}_{AB} \Bigr), 
\label{7.4}
\end{equation} 
where $\Psi^{lm}_{\rm even} \equiv \Psi^{lm}_{\rm even}(v,r=2M)$ 
and $\Psi^{lm}_{\rm odd} \equiv \Psi^{lm}_{\rm odd}(v,r=2M)$. This
result should be compared with Eq.~(\ref{6.13}). The rates at which
the gravitational perturbation transfers energy and angular momentum
to the black hole can now be calculated using the method described in
Sec.~VII of Poisson \cite{poisson:04c}. Our Eq.~(\ref{7.4}) replaces
his Eq.~(7.5), and the rest of the calculation is identical. The final
results are   
\begin{eqnarray}
\biggl\langle \frac{dE}{dv} \biggr\rangle &=& \frac{1}{64\pi} 
\sum_{lm} (l-1)l(l+1)(l+2) 
\nonumber \\ & & \mbox{} \times 
\Bigl\langle \bigl| \dot{\Psi}^{lm}_{\rm even} \bigr|^2
+ \bigl| \dot{\Psi}^{lm}_{\rm odd} \bigr|^2 \Bigr\rangle 
\label{7.5}
\end{eqnarray}
and
\begin{eqnarray}
\biggl\langle \frac{dJ}{dv} \biggr\rangle &=& \frac{1}{64\pi} 
\sum_{lm} (l-1)l(l+1)(l+2) (im) 
\nonumber \\ & & \mbox{} \times 
\Bigl\langle \bar{\Psi}^{lm}_{\rm even} \dot{\Psi}^{lm}_{\rm even} 
+ \bar{\Psi}^{lm}_{\rm odd} \dot{\Psi}^{lm}_{\rm odd} \Bigr\rangle.   
\label{7.6}
\end{eqnarray}
These equations replace Poisson's Eqs.~(7.8) and (7.9). Notice the
similarity between Eqs.~(\ref{7.5}) and (\ref{6.16}), and between 
Eqs.~(\ref{7.6}) and (\ref{6.17}). In Eq.~(\ref{7.6}), $J$ represents  
the component of the hole's angular-momentum vector in the arbitrary
$z$ direction which defines the orientation of the angles $\theta$ and 
$\phi$. The overbar indicates complex conjugation. The averaging
carried out in Eqs.~(\ref{7.5}) and (\ref{7.6}) is over a
characteristic time scale associated with the gravitational
perturbation. 

\section{Low multipoles} 
\label{viii} 

To conclude our presentation of the perturbation formalism we now 
handle the special cases $l = 0$ and $l = 1$. Our discussion here will
essentially reproduce Zerilli's Appendix G \cite{zerilli:70}, but we
will frame it in the language developed in this paper. For
concreteness, and for simplicity, we will focus mostly on the task of
integrating the {\it vacuum} field equations for these low multipoles.   

\subsection{Monopole perturbation} 

When $l=0$ the only relevant spherical harmonic is $Y^{00}$, which is
a constant. It follows that $Y_A = X_A = Y_{AB} = X_{AB} = 0$, and the
perturbation is of even parity. The only nonvanishing metric
perturbations are   
\begin{equation} 
p_{ab} = h_{ab} Y^{00}, \qquad
p_{AB} = r^2 K \Omega_{AB} Y^{00}; 
\label{8.1}
\end{equation}
the fields $h_a$ and $G$ are not defined. The freedom to perform a 
gauge transformation is contained in $\Xi_a = \xi_a Y^{00}$, $\Xi_A =
0$, and the perturbations transform as   
\begin{eqnarray}
h_{ab} &\to& h'_{ab} = h_{ab} - \nabla_a \xi_b - \nabla_b \xi_a, 
\label{8.2} \\ 
K &\to& K' = K - \frac{2}{r} r^a \xi_a. 
\label{8.3}
\end{eqnarray} 
There is no analogue here of the gauge-invariant quantities
$\tilde{h}_{ab}$ and $\tilde{K}$ that were introduced in the general    
case. The relevant field equations for $h_{ab}$ and $K$ are the
$Q_{ab}$ and $Q^\flat$ equations of Eqs.~(\ref{4.13}) and
(\ref{4.15}), respectively, in which we set $l=0$, $\tilde{h}_{ab}  
= h_{ab}$, and $\tilde{K} = K$. The $Q_a$ and $Q^\sharp$ equations of 
Eqs.~(\ref{4.14}) and (\ref{4.16}), respectively, are not defined. 

We now solve the perturbation equations in the absence of material
sources; in this case $Q_{ab} = 0 = Q^\flat$. We adopt the advanced
coordinates $(v,r,\theta,\phi)$ and the metric of Eq.~(\ref{2.4}). For
convenience we rescale the value of $Y^{00}$ and set it equal to
unity. We choose $\xi_a$ so as to enforce the gauge conditions $h_{rr}
= K = 0$. This does not fully determine the gauge, because these
conditions are preserved under an additional transformation generated
by $\xi_v = -f \alpha(v)$ and $\xi_r = \alpha(v)$, where $f = 1-2M/r$
and $\alpha(v)$ is an arbitrary function. This transformation
corresponds to a redefinition of the advanced-time coordinate $v$, of
the form $v \to v' = v + \xi^v = v + \alpha(v)$, and it changes the
nonvanishing perturbations according to $h_{vv} \to h'_{vv} = h_{vv} 
+ 2 f \dot{\alpha}(v)$ and $h_{vr} \to h'_{vr} = h_{vr} 
- \dot{\alpha}(v)$, where an overdot indicates differentiation with
respect to $v$.   

We now involve the field equations. We notice first that the $Q_{rr} =
0$ equation implies that $h_{vr}$ is in fact independent of $r$; since 
this function of $v$ can be altered at will by a gauge transformation,
we may set $h_{vr} = 0$. The gauge is now fully determined, and the
sole remaining perturbation field is $h_{vv}$. By virtue of the
$Q_{vr} = 0$ equation we have that $r h_{vv}$ must be a function of
$v$ only, but the $Q_{vv} = 0$ equation constrains this function to be
a constant. We therefore have $h_{vv} = (\mbox{constant})/r$, and this 
completes the integration of the perturbation equations.  

The vacuum, monopole perturbation of a Schwarzschild black hole is
given by  
\begin{equation} 
p_{vv} = \frac{2\delta M}{r},
\label{8.4}
\end{equation} 
where $\delta M$ is a constant; all other components vanish. This
addition to $g_{vv}$ simply shifts the Schwarzschild mass parameter
from $M$ to $M + \delta M$. The perturbation must respect Birkhoff's
theorem, and the perturbed metric is indeed another Schwarzschild
solution.   

\subsection{Odd-parity dipole perturbation} 

The only surviving odd-parity spherical harmonics for $l=1$ are
$X_A^{1m}$, which are obtained from $Y^{1m}$ using Eq.~(\ref{3.2}). 
The tensorial harmonics $X_{AB}^{1m}$ vanish, and the only surviving
components of the metric perturbation are 
\begin{equation} 
p_{aB} = \sum_{m} h^{1m}_a X_B^{1m}. 
\label{8.5}
\end{equation}
The perturbations $h^{1m}_a$ can be altered by a gauge transformation 
generated by $\Xi_a = 0$, $\Xi_A = \sum_{m} \xi^{1m} X_A^{1m}$; they 
change according to 
\begin{equation}
h_a \to h'_a = h_a - \nabla_a \xi + \frac{2}{r} r_a \xi. 
\label{8.6}
\end{equation} 
There is no analogue here of the gauge-invariant fields $\tilde{h}_a$
introduced in Sec.~\ref{v} for the general case. The relevant field
equation for $h_{a}$ is the $P_{a}$ equation of Eq.~(\ref{5.8}), in
which we set $l=1$ and $\tilde{h}_{a} = h_{a}$. The $P$ equation of
Eq.~(\ref{5.9}) is not defined. 
 
We now solve the perturbation equations in the absence of material
sources; in this case $P_{a} = 0$. We once more adopt the advanced
coordinates $(v,r,\theta,\phi)$ and the metric of Eq.~(\ref{2.4}). For 
convenience we replace the set of complex harmonics $Y^{1m}$ with the
set $Y^{\sf m}$ given by  
\begin{eqnarray} 
(Y^0, Y^{s}, Y^{c})  
&=& (\cos\theta,\sin\theta\sin\phi,\sin\theta\cos\phi) 
\nonumber \\ 
&=:& (\Omega^3, \Omega^2, \Omega^1).
\label{8.7}
\end{eqnarray}  
This set is real, but not normalized; summation over $m$ will be 
replaced by a summation over the abstract index ${\sf m} = (0,s,c)$.  
Notice that $\Omega^i$ represents the Cartesian components of a unit
radial vector; in ordinary vectorial notation, $\bm{\Omega} =
\bm{r}/r$, where $\bm{r} = (x,y,z)$ and $r = |\bm{r}|$. We shall also
need $\Omega^i_A := \partial \Omega^i/\partial \theta^A$, the
derivatives of $\Omega^i$ with respect to the angles $\theta^A$.      

We choose $\xi$ so as to enforce the gauge condition $h_{r} =
0$. This does not fully determine the gauge, because this condition is 
preserved under an additional transformation generated by $\xi =
\alpha(v) r^2$, where $\alpha(v)$ is an arbitrary function. This
transformation changes $h_v$ according to $h_{v} \to h'_{v} = h_{v} -
\dot{\alpha}(v) r^2$, where an overdot indicates differentiation with
respect to $v$.    

We now involve the field equations. We note first that the $P_{r} = 0$  
equation implies that $h_v$ must have the form of $h_v = k_1(v) r^2 +
k_2(v)/r$, where $k_1$ and $k_2$ are arbitrary functions. The first
term can be removed by a gauge transformation, and we are left with
$h_v = k_2(v)/r$ and a fully determined gauge. The $P_v = 0$ equation
then implies that $k_2$ must in fact be a constant. We therefore have
$h_v = (\mbox{constant})/r$, and this completes the integration of the
perturbation equations. 

The vacuum, odd-parity, dipole perturbation of a Schwarzschild black
hole is given by  
\[
p_{vA} = \frac{2}{r} \sum_{\sf m} (\delta J)_{\sf m} X^{\sf m}_A, 
\]
where $(\delta J)_{\sf m}$ are the constants identified in the
preceding paragraph, and $X^{\sf m}_A = -\varepsilon_A^{\ B} D_B 
Y^{\sf m}$ are the real odd-parity harmonics; all other components of
the metric perturbation vanish. Relating the spherical harmonics to
the unit radial vector $\Omega^i$ as in Eq.~(\ref{8.7}), and writing
the summation over ${\sf m}$ as an implicit summation over $i$, we
obtain our final expression,  
\begin{equation} 
p_{vA} = -\frac{2}{r} \varepsilon_A^{\ B} (\delta J)_i \Omega^i_B. 
\label{8.8}
\end{equation} 
To give an interpretation to this result we perform a transformation
from the spherical coordinates $(r,\theta^A)$ to the Cartesian
coordinates $x^i = r \Omega^i$. The metric perturbation transforms as
$p_{vi} = r^{-1} p_{vA} \Omega^A_i$, where $\Omega^A_i := \Omega^{AB}
\delta_{ij} \Omega^j_B$. This gives $p_{vi} = -2r^{-2}
\varepsilon_{AB} \Omega^A_{i} \Omega^B_{j} (\delta J)^j$. We now
involve the easily-established identity $\varepsilon_{AB} \Omega^A_{i} 
\Omega^B_{j} = \varepsilon_{ijk} \Omega^k$, where $\varepsilon_{ijk}$
is the permutation symbol (a fully antisymmetric tensor with entries
$-1$, $0$, or $1$). This gives 
\begin{equation}
p_{vi} = -\frac{2}{r^2} \varepsilon_{ijk} (\delta J)^j \Omega^k 
= - \frac{2}{r^3} (\bm{\delta J} \times \bm{r})_i. 
\label{8.9}
\end{equation} 
This form for the metric tensor allows us to identify $\bm{\delta J}$
with the angular-momentum vector of the spacetime. The perturbed
metric therefore describes a slowly rotating black hole; it is the
Kerr metric linearized with respect to its angular-momentum
parameter. 

\subsection{Even-parity dipole perturbation} 

The even-parity spherical harmonics for $l=1$ are $Y^{1m}$,
$Y^{1m}_A$, and $\Omega_{AB} Y^{1m}$; $Y_{AB}^{1m}$ vanishes. The
metric perturbation is then 
\begin{eqnarray} 
p_{ab} &=& \sum_{m} h^{1m}_{ab} Y^{1m}, 
\label{8.10} \\ 
p_{aB} &=& \sum_{m} j^{1m}_a Y^{1m}_B, 
\label{8.11} \\ 
p_{AB} &=& r^2 \Omega_{AB} \sum_{m} K^{1m} Y^{1m}, 
\label{8.12} 
\end{eqnarray}
and the fields $G^{1m}$ are not defined. Gauge transformations are
generated by $\Xi_a = \sum_m \xi^{1m}_a Y^{1m}$ and $\Xi_A = \sum_m
\xi^{1m} Y^{1m}_A$. The perturbations change according to  
\begin{eqnarray}
h_{ab} &\to& h'_{ab} = h_{ab} - \nabla_a \xi_b - \nabla_b \xi_a, 
\label{8.13} \\ 
j_{a} &\to& j'_{a} = j_a - \xi_a -\nabla_a \xi + \frac{2}{r} r_a \xi,  
\label{8.14} \\ 
K &\to& K' = K + \frac{2}{r^2} \xi - \frac{2}{r} r^a \xi_a. 
\label{8.15} 
\end{eqnarray} 
There is no analogue here of the gauge-invariant fields 
$\tilde{h}_{ab}$ and $\tilde{K}$ introduced in Sec.~\ref{iv} for the
general case. The relevant field equations for $h_{ab}$, $j_a$, and
$K$ are the $Q_{ab}$, $Q_a$, $Q^\flat$ equations of
Eqs.~(\ref{4.13})--(\ref{4.15}), in which we set $l=1$,
$\tilde{h}_{ab} = h_{ab} - \nabla_a j_b - \nabla_b j_a$, and 
$\tilde{K} = K - 2 r^{-1} r^a j_a$. The $Q^\sharp$ equation of
Eq.~(\ref{4.16}) is not defined for $l=1$.    
 
We now solve the vacuum field equations, $Q_{ab} = Q_a = Q^\flat =
0$. We adopt the advanced coordinates $(v,r,\theta,\phi)$, the real 
spherical harmonics of Eq.~(\ref{8.7}), and a gauge in which $h_{rr} =
h_r = K = 0$. This gauge is preserved under a transformation generated
by    
\begin{eqnarray*}  
\xi_v &=& \frac{2M}{r} \alpha_1(v) + \alpha_2(v) r, \\ 
\xi_r &=& \alpha_1(v), \\ 
\xi &=& \alpha_1(v) r + \alpha_2(v) r^2,  
\end{eqnarray*} 
where $\alpha_1(v)$ and $\alpha_2(v)$ are arbitrary functions. This
has the following effect on the remaining components of the metric
perturbation:   
\begin{eqnarray*} 
h_{vv} &\to& h'_{vv} = h_{vv} - 2 \dot{\alpha}_2 r - \frac{2M}{r} 
(2 \dot{\alpha}_1 - \alpha_2) + \frac{2M}{r^2} \alpha_1, \\ 
h_{vr} &\to& h'_{vr} = h_{vr} - (\dot{\alpha}_1 + \alpha_2), \\  
j_v &\to& j'_v = j_v - \dot{\alpha}_2 r^2 - (\dot{\alpha}_1 
+ \alpha_2) r - \frac{2M}{r} \alpha_1. 
\end{eqnarray*} 
The $Q_{rr} = 0$ equation implies that $h_{vr}$ is a function of $v$ 
only, and this component can be set equal to zero by a gauge 
transformation. The remaining gauge freedom is now restricted by  
$\alpha_2 = -\dot{\alpha}_1$. The $Q_{r} = 0$ equation implies
that $j_v$ must be of the form $j_v = k_1(v) r^2 + k_2(v)/r$, where
$k_1$ and $k_2$ are arbitrary functions of $v$. With this information
and the $Q_v = 0$ equation we find that $h_{vv}$ must of the form
$h_{vv} = -k_2/r^2 + 3\dot{k}_2/r + k_3 + 2k_1 r$, where $k_3$ is an
additional arbitrary function. The $Q_{vr} = 0$ equation now reveals
that $k_3 = 0$, the $Q_{vv} = 0$ equation yields $2M k_1 =
-\ddot{k}_2$, and the task of integrating the perturbation equations
is completed. Defining $k(v) := k_2(v)/(2M)$, we have obtained   
\[
h_{vv} = -2 \ddot{k}(v) r + \frac{6M}{r} \dot{k}(v) 
- \frac{2M}{r^2} k(v)
\]
and
\[
j_v = -\ddot{k}(v) r^2 + \frac{2M}{r} k(v) 
\]
for the nonvanishing perturbation fields. The remaining gauge freedom
is contained in  
\begin{eqnarray*}  
\xi_v &=& -\dot{\alpha}(v) r + \frac{2M}{r} \alpha(v), \\  
\xi_r &=& \alpha(v), \\ 
\xi &=& -\dot{\alpha}(v) r^2 + \alpha(v) r,  
\end{eqnarray*} 
where we have set $\alpha := \alpha_1$ and $\alpha_2 =
-\dot{\alpha}$. Under this transformation the perturbations change 
according to 
\[
h_{vv} \to h'_{vv} = h_{vv} + 2 \ddot{\alpha}(v) r - \frac{6M}{r}
\dot{\alpha}(v) + \frac{2M}{r^2} \alpha(v)
\]
and 
\[
j_v \to j'_v = j_v + \ddot{\alpha}(v) r^2 - \frac{2M}{r} \alpha(v).  
\]
Setting $\alpha(v) = k(v)$ produces $h'_{vv} = j'_v = 0$, and we
conclude that the perturbation is pure gauge. 
 
The vacuum, even-parity, dipole perturbation of a Schwarzschild black 
hole can be presented in a gauge in which the nonvanishing components
are  
\begin{equation} 
p_{vv} = -2 r \ddot{\alpha}_i \Omega^i  + \frac{6M}{r}
\dot{\alpha}_i \Omega^i - \frac{2M}{r^2} \alpha_i \Omega^i
\label{8.16} 
\end{equation}
and 
\begin{equation} 
p_{vA} = -r^2 \ddot{\alpha}_i \Omega^i_A + \frac{2M}{r} \alpha_i
\Omega^i_A, 
\label{8.17}
\end{equation} 
where the vector $\bm{\alpha}(v)$ is an arbitrary function of the
advanced-time coordinate $v$. This perturbation, however, can be
removed by a gauge transformation generated by  
\begin{eqnarray} 
\Xi_v &=& -r \dot{\alpha}_i \Omega^i 
          + \frac{2M}{r} \alpha_i \Omega^i,
\label{8.18} \\ 
\Xi_r &=& \alpha_i \Omega^i, 
\label{8.19} \\ 
\Xi_A &=& -r^2 \dot{\alpha}_i \Omega^i_A + r \alpha_i \Omega^i_A. 
\label{8.20}
\end{eqnarray} 
The perturbed metric is therefore a Schwarzschild solution expressed
in a coordinate system which differs from the original system
$(v,r,\theta^A)$. These coordinates are not inertial. In the limit 
$M \to 0$ the perturbed metric is recognized as the metric of flat 
spacetime expressed in a light-cone coordinate system centered on an
accelerated world line (see Ref.~\cite{poisson:04a}); the acceleration
vector is $\ddot{\alpha}_i(v)$. For $M \neq 0$ the perturbed metric
describes a Schwarzschild black hole moving on ``the same''
accelerated world line.   

\begin{acknowledgments}
This work was supported by the National Science and Engineering
Research Council of Canada. 
\end{acknowledgments}

\appendix
\section{Components of the vector and tensor spherical harmonics, and
relationship with other spherical harmonics} 
\label{A}

We first list the components of the tensorial harmonics introduced in
Sec.~\ref{iii}. According to Eq.~(\ref{3.1}) we have 
\begin{eqnarray*} 
Y^{lm}_\theta &=& \frac{\partial}{\partial \theta} Y^{lm}, \\ 
Y^{lm}_\phi &=& \frac{\partial}{\partial \phi} Y^{lm}.
\end{eqnarray*} 
According to Eq.~(\ref{3.2}) we have 
\begin{eqnarray*} 
X^{lm}_\theta &=& -\frac{1}{\sin\theta} 
\frac{\partial}{\partial \phi} Y^{lm}, \\ 
X^{lm}_\phi &=& \sin\theta \frac{\partial}{\partial \theta} Y^{lm}. 
\end{eqnarray*} 
According to Eq.~(\ref{3.6}) we have 
\begin{eqnarray*} 
Y^{lm}_{\theta\theta} &=& \biggl[ \frac{\partial^2}{\partial \theta^2} 
+ \frac{1}{2} l(l+1) \biggr] Y^{lm}, \\  
Y^{lm}_{\theta\phi} &=& \biggl[ \frac{\partial^2}{\partial \theta
\partial \phi} - \frac{\cos\theta}{\sin\theta}
  \frac{\partial}{\partial \phi} \biggr] Y^{lm}, \\ 
Y^{lm}_{\phi\phi} &=& \biggl[ \frac{\partial^2}{\partial \phi^2} 
+ \sin\theta\cos\theta \frac{\partial}{\partial \theta} 
+ \frac{1}{2} l(l+1) \sin^2\theta \biggr] Y^{lm}.
\end{eqnarray*} 
And according to Eq.~(\ref{3.7}) we have 
\begin{eqnarray*} 
X^{lm}_{\theta\theta} &=& -\frac{1}{\sin\theta} \biggl( 
\frac{\partial^2}{\partial \theta \partial \phi} 
- \frac{\cos\theta}{\sin\theta} \frac{\partial}{\partial \phi} \biggr)
Y^{lm}, \\
X^{lm}_{\theta\phi} &=& \frac{1}{2} \biggl( \sin\theta  
\frac{\partial^2}{\partial \theta^2} - \frac{1}{\sin\theta} 
\frac{\partial^2}{\partial \phi^2} - \cos\theta
\frac{\partial}{\partial \theta} \biggr) Y^{lm}, \\ 
X^{lm}_{\phi\phi} &=& \biggl( \sin\theta 
\frac{\partial^2}{\partial \theta \partial \phi} 
- \cos\theta \frac{\partial}{\partial \phi} \biggr) Y^{lm}.
\end{eqnarray*} 

The tensorial harmonics $Y_{AB}^{lm}$ and $X_{AB}^{lm}$ can be related 
to the spherical-harmonic functions of spin-weight $s=\pm 2$
\cite{goldberg-etal:67}. Let $m_A$ and $\bar{m}_A$ be a complex
orthonormal basis on the unit two-sphere, with 
$m_A = 2^{-1/2} (1,i\sin\theta)$. The relationship is then 
\begin{eqnarray*} 
Y_{AB}^{lm} &=& \frac{1}{2} \sqrt{(l-1)l(l+1)(l+2)} 
\\ & & \mbox{} \times 
\bigl( \mbox{}_{-2} Y^{lm} m_A m_B 
+ \mbox{}_2 Y^{lm} \bar{m}_A \bar{m}_B \bigr) 
\end{eqnarray*} 
and 
\begin{eqnarray*} 
X_{AB}^{lm} &=& -\frac{i}{2} \sqrt{(l-1)l(l+1)(l+2)}
\\ & & \mbox{} \times 
\bigl( \mbox{}_{-2} Y^{lm} m_A m_B 
- \mbox{}_2 Y^{lm} \bar{m}_A \bar{m}_B \bigr),   
\end{eqnarray*} 
where $\mbox{}_s Y^{lm}$ are the spin-weighted spherical harmonics. 
These equations can be compared with Eqs.~(2.38e) and (2.38f) of 
Ref.~\cite{thorne:80}. This reveals that our tensorial harmonics are 
intimately related to the ``pure-spin'' harmonics used by Thorne. The 
relationship is 
\[ 
Y_{AB}^{lm} = \sqrt{{\textstyle \frac{1}{2}} (l-1)l(l+1)(l+2)}\, 
T^{E2,lm}_{AB} 
\]
and 
\[
X_{AB}^{lm} = \sqrt{{\textstyle \frac{1}{2}} (l-1)l(l+1)(l+2)}\, 
T^{B2,lm}_{AB}.  
\]
Notice that the pure-spin harmonics $T^{E2,lm}_{AB}$ and 
$T^{B2,lm}_{AB}$ are normalized on the unit two-sphere. Our
convention here differs from Thorne's, who inserts a factor of
$r^{-1}$ in $m_A$ and $\bar{m}_A$ in order to normalize them on a
two-sphere of radius $r$.  

\section{Perturbation of the Ricci tensor for a general
spherically-symmetric background spacetime} 
\label{B} 

The perturbed spacetime metric is written as 
\begin{eqnarray*} 
\mbox{}^4 g_{ab} &=& g_{ab} + p_{ab}, \\ 
\mbox{}^4 g_{aB} &=& p_{aB}, \\ 
\mbox{}^4 g_{AB} &=& r^2 \Omega_{AB} + p_{AB}, 
\end{eqnarray*} 
where $(g_{ab},r^2\Omega_{AB})$ are the components of the background
metric and $(p_{ab},p_{aB},p_{AB})$ are the components of the
perturbation. In this Appendix we allow the background metric to be
completely general, so long as it is spherically symmetric; we do not
restrict it to be the Schwarzschild metric. We raise lower-case Latin
indices with $g^{ab}$, the inverse to $g_{ab}$, and we raise
upper-case Latin indices with $\Omega^{AB}$, the inverse to
$\Omega_{AB}$. The inverse perturbed metric is thus 
\begin{eqnarray*} 
\mbox{}^4 g^{ab} &=& g^{ab} - p^{ab}, \\ 
\mbox{}^4 g^{aB} &=& -\frac{1}{r^2} p^{aB}, \\ 
\mbox{}^4 g^{AB} &=& \frac{1}{r^2} \Omega^{AB} - \frac{1}{r^4} p^{AB},    
\end{eqnarray*} 
up to terms quadratic in the perturbations. 

Covariant differentiation with respect to the coordinates $x^a$ on the
submanifold ${\cal M}^2$ is indicated with $\nabla_a$: 
$\nabla_a g_{bc} = 0$. Covariant differentiation with respect to the
coordinates $\theta^A$ on the unit two-sphere is indicated with $D_A$:
$D_A \Omega_{BC} = 0$. Quantities which depend only on $x^a$ are
covariantly constant relative to the connection $\Gamma^A_{\ BC}$; for
example $D_A r \equiv 0$. Quantities which depend only on $\theta^A$
are covariantly constant relative to the connection $\Gamma^a_{\ bc}$;
for example $\nabla_a \Omega_{AB} \equiv 0$.   

A straightforward calculation returns the components of the perturbed
connection, which we denote $\mbox{}^4 \Gamma^\alpha_{\ \beta\gamma} +
\delta \Gamma^\alpha_{\ \beta\gamma}$. We obtain    
\begin{eqnarray*}
\delta \Gamma^a_{\ bc} &=& C^a_{\ bc}, \\ 
\delta \Gamma^a_{\ bC} &=& \frac{1}{2} 
\bigl( D_C p^a_{\ b} + \nabla_b p^a_{\ C} - \nabla^a p_{bC} \bigr)  
- \frac{1}{r} r_b p^a_{\ C}, \\ 
\delta \Gamma^a_{\ BC} &=& \frac{1}{2} 
\bigl( D_B p^a_{\ C} + D_C p^a_{\ B} - \nabla^a p_{BC} \bigr) 
+ r r_m \Omega_{BC} p^{am}, \\  
\delta \Gamma^A_{\ bc} &=& \frac{1}{2r^2}  
\bigl( \nabla_b p_{c}^{\ A} + \nabla_c p_{b}^{\ A} - D^A p_{bc}
\bigr), \\   
\delta \Gamma^A_{\ bC} &=& \frac{1}{2r^2} 
\bigl( D_C p_{b}^{\ A} - D^A p_{bC} + \nabla_b p^A_{\ C} \bigr) 
- \frac{1}{r^3} r_b p^A_{\ C}, \\ 
\delta \Gamma^A_{\ BC} &=& \frac{1}{r^2} C^A_{\ BC} +   
\frac{1}{r} r_m \Omega_{BC} p^{mA}, 
\end{eqnarray*} 
where $r_a := \nabla_a r$, 
\[
C^a_{\ bc} := \frac{1}{2} 
\bigl( \nabla_c p^a_{\ b} + \nabla_b p^a_{\ c} - \nabla^a p_{bc}
\bigr),  
\]
and 
\[
C^A_{\ BC} := \frac{1}{2} 
\bigl( D_C p^A_{\ B} + D_B p^A_{\ C} - D^A p_{BC} \bigr). 
\]

The Ricci tensor for the perturbed spacetime is equal to the
background Ricci tensor $R_{\alpha\beta}$ plus its perturbation 
$\delta R_{\alpha\beta}$. We obtain 
\begin{widetext} 
\begin{eqnarray*} 
\delta R_{ab} &=& \nabla_m C^m_{\ ab} + \frac{2}{r} r_m C^m_{\ ab}  
- \frac{1}{2} \nabla_a \nabla_b p^m_{\ m} 
- \frac{1}{2r^2} D^M D_M p_{ab} 
+ \frac{1}{2r^2} D_M \bigl( \nabla_a p_b^{\ M} 
                              + \nabla_b p_a^{\ M} \bigr)
\\ & & \mbox{} 
- \frac{1}{2r^2} \nabla_a \nabla_b p^M_{\ M} 
+ \frac{1}{2r^3} \bigl( r_a \nabla_b p^M_{\ M} 
                          + r_b \nabla_a p^M_{\ M} \bigr) 
- \frac{1}{r^4} (r_a r_b - r \nabla_a \nabla_b r) p^M_{\ M}, \\ 
\delta R_{aB} &=& \frac{1}{2} D_B \Bigl( \nabla_m p^m_{\ a} 
- \nabla_a p^m_{\ m} + \frac{1}{r} r_a p^m_{\ m} \Bigr) 
- \frac{1}{2} \bigl( \Box p_{aB} - \nabla_m \nabla_a p^m_{\ B} \bigr) 
- \frac{1}{r} \bigl( r_a \nabla_m p^m_{\ B} 
                      - r_m \nabla_a p^m_{\ B} \bigr)   
\\ & & \mbox{} 
- \frac{1}{r^2} \bigl( r_a r_m 
                       + r \nabla_a \nabla_m r \bigr) p^m_{\ B}  
+ \frac{1}{2r^2} D^M \bigl( D_B p_{aM} - D_M p_{aB} \bigr) 
+ \frac{1}{2r^2} \nabla_a \bigl( D_M p^M_{\ B} 
                                   - D_B p^M_{\ M} \bigr)
\\ & & \mbox{} 
- \frac{1}{r^3} r_a \bigl( D_M p^M_{\ B} - D_B p^M_{\ M} \bigr), \\ 
\delta R_{AB} &=& \Omega_{AB} \biggl[ r r_m \nabla_n \Bigl( p^{mn} 
- \frac{1}{2} g^{mn} p^k_{\ k} \Bigr) 
+ (r_m r_n + r \nabla_m \nabla_n r) p^{mn} \biggr] 
- \frac{1}{2} D_A D_B p^m_{\ m} 
+ \frac{1}{2} \nabla_m \bigl( D_A p^m_{\ B} + D_B p^m_{\ A} \bigr) 
\\ & & \mbox{} 
+ \frac{1}{r} r_m \Omega_{AB} D_M p^{mM}  
- \frac{1}{2} \Box p_{AB} + \frac{1}{r^2} D_M C^M_{\ AB} 
- \frac{1}{2r^2} D_A D_B p^M_{\ M} 
+ \frac{1}{r} r^m \nabla_m \Bigl( p_{AB} 
             - \frac{1}{2} \Omega_{AB} p^M_{\ M} \Bigr)           
\\ & & \mbox{} 
- \frac{2}{r^2} r^m r_m \Bigl( p_{AB} 
             - \frac{1}{2} \Omega_{AB} p^M_{\ M} \Bigr),  
\end{eqnarray*}     
where $\Box := g^{ab} \nabla_a \nabla_b$. These expressions can be
simplified by involving Eqs.~(\ref{2.6}) and (\ref{2.8}) when the
background spacetime is the Schwarzschild spacetime.  
 
Under a gauge transformation generated by the dual vector field
$\Xi_\alpha = (\Xi_a,\Xi_A)$, the components of the metric
perturbation change according to 
\begin{eqnarray*} 
p_{ab} &\to& p'_{ab} := p_{ab} - \nabla_a \Xi_b - \nabla_b \Xi_a, \\  
p_{aB} &\to& p'_{aB} := p_{aB} - \nabla_a \Xi_B - D_B \Xi_a 
             + \frac{2}{r} r_a \Xi_B, \\ 
p_{AB} &\to& p'_{AB} := p_{AB} - D_A \Xi_B - D_B \Xi_A 
             - 2 r r^m \Xi_m \Omega_{AB}.  
\end{eqnarray*} 
It can be shown that when the background Ricci tensor vanishes, 
$\delta R_{ab}$, $\delta R_{aB}$, and $\delta R_{AB}$ are all
invariant under this transformation. We use this property in
Secs.~\ref{iv} and \ref{v}.      

\section{Perturbation equations in $(t,r)$ coordinates} 
\label{C}

In this Appendix the two-dimensional Schwarzschild metric is written
in its standard form 
\[
ds^2 = -f\, dt^2 + f^{-1}\, dr^2, 
\]
with $f = 1-2M/r$. In these coordinates the wave operator is given by 
\[
\Box \Psi = \biggl( -\frac{1}{f} \frac{\partial^2}{\partial t^2}  
+ \frac{\partial}{\partial r} f \frac{\partial}{\partial r} \biggr) 
\Psi. 
\]
We set $\lambda := l(l+1) = \mu + 2$ and $\mu := (l-1)(l+2) 
= \lambda - 2$. 

The even-parity perturbations are $h_{tt}$, $h_{tr}$, $h_{rr}$, $j_t$,
$j_r$, $K$, and $G$. Under a gauge transformation generated by
$\xi_t$, $\xi_r$, and $\xi$ they change according to $h_{tt} \to
h'_{tt} = h_{tt} + \Delta h_{tt}$, etc., with   
\begin{eqnarray*} 
\Delta h_{tt} &=& -2 \frac{\partial}{\partial t} \xi_t 
+ \frac{2Mf}{r^2} \xi_r, \\ 
\Delta h_{tr} &=& -\frac{\partial}{\partial r} \xi_t 
- \frac{\partial}{\partial t} \xi_r + \frac{2M}{r^2 f} \xi_t, \\
\Delta h_{rr} &=& -2 \frac{\partial}{\partial r} \xi_r 
- \frac{2M}{r^2f} \xi_r, \\
\Delta j_t &=& -\frac{\partial}{\partial t} \xi - \xi_t, \\ 
\Delta j_r &=& -\frac{\partial}{\partial r} \xi - \xi_r 
+ \frac{2}{r} \xi, \\
\Delta K &=& -\frac{2f}{r} \xi_r + \frac{\lambda}{r^2} \xi, \\ 
\Delta G &=& -\frac{2}{r^2} \xi. 
\end{eqnarray*}     
The gauge-invariant fields are 
\begin{eqnarray*} 
\tilde{h}_{tt} &=& h_{tt} - 2 \frac{\partial}{\partial t} j_t 
+ \frac{2Mf}{r^2} j_r + r^2 \frac{\partial^2}{\partial t^2} G 
- Mf \frac{\partial}{\partial r} G, \\
\tilde{h}_{tr} &=& h_{tr} - \frac{\partial}{\partial r} j_t 
- \frac{\partial}{\partial t} j_r + \frac{2M}{r^2 f} j_t
+ r^2 \frac{\partial^2}{\partial t \partial r} G 
+ \frac{r-3M}{f} \frac{\partial}{\partial t} G, \\ 
\tilde{h}_{rr} &=& h_{rr} - 2 \frac{\partial}{\partial r} j_r 
- \frac{2M}{r^2 f} j_r + r^2 \frac{\partial^2}{\partial r^2} G 
+ \frac{2r-3M}{f} \frac{\partial}{\partial r} G, \\
\tilde{K} &=& K - \frac{2f}{r} j_r 
+ rf \frac{\partial}{\partial r} G
+ \frac{\lambda}{2} G. 
\end{eqnarray*} 
The field equations are
\begin{eqnarray*} 
Q^{tt} &=& -\frac{\partial^2}{\partial r^2} \tilde{K} 
- \frac{3r-5M}{r^2 f} \frac{\partial}{\partial r} \tilde{K} 
+ \frac{f}{r} \frac{\partial}{\partial r} \tilde{h}_{rr}
+ \frac{(\lambda+2)r + 4M}{2r^3} \tilde{h}_{rr} 
+ \frac{\mu}{2r^2 f} \tilde{K}, \\ 
Q^{tr} &=& \frac{\partial^2}{\partial t \partial r} \tilde{K} 
+ \frac{r-3M}{r^2 f} \frac{\partial}{\partial t} \tilde{K} 
- \frac{f}{r} \frac{\partial}{\partial t} \tilde{h}_{rr} 
- \frac{\lambda}{2r^2} \tilde{h}_{tr}, \\
Q^{rr} &=& -\frac{\partial^2}{\partial t^2} \tilde{K} 
+ \frac{(r-M)f}{r^2} \frac{\partial}{\partial r} \tilde{K} 
+ \frac{2f}{r} \frac{\partial}{\partial t} \tilde{h}_{tr} 
- \frac{f}{r} \frac{\partial}{\partial r} \tilde{h}_{tt}  
+ \frac{\lambda r + 4M}{2r^3} \tilde{h}_{tt} 
- \frac{f^2}{r^2} \tilde{h}_{rr} 
- \frac{\mu f}{2r^2} \tilde{K}, \\ 
Q^{t} &=& \frac{\partial}{\partial t} \tilde{h}_{rr}
- \frac{\partial}{\partial r} \tilde{h}_{tr} 
+ \frac{1}{f} \frac{\partial}{\partial t} \tilde{K} 
- \frac{2M}{r^2 f} \tilde{h}_{tr}, \\
Q^{r} &=& -\frac{\partial}{\partial t} \tilde{h}_{tr}
+ \frac{\partial}{\partial r} \tilde{h}_{tt} 
- f \frac{\partial}{\partial r} \tilde{K} 
- \frac{r-M}{r^2 f} \tilde{h}_{tt}
+ \frac{(r-M)f}{r^2} \tilde{h}_{rr}, \\
Q^\flat &=& -\frac{\partial^2}{\partial t^2} \tilde{h}_{rr} 
+ 2 \frac{\partial^2}{\partial t \partial r} \tilde{h}_{tr} 
- \frac{\partial^2}{\partial r^2} \tilde{h}_{tt} 
- \frac{1}{f} \frac{\partial^2}{\partial t^2} \tilde{K} 
+ f \frac{\partial^2}{\partial r^2} \tilde{K} 
+ \frac{2(r-M)}{r^2 f} \frac{\partial}{\partial t} \tilde{h}_{tr} 
- \frac{r-3M}{r^2 f} \frac{\partial}{\partial r} \tilde{h}_{tt} 
\\ & & \mbox{} 
- \frac{(r-M)f}{r^2} \frac{\partial}{\partial r} \tilde{h}_{rr} 
+ \frac{2(r-M)}{r^2} \frac{\partial}{\partial r} \tilde{K} 
+ \frac{\lambda r^2-2(2+\lambda)Mr+4M^2}{2r^4 f^2} \tilde{h}_{tt}   
- \frac{\lambda r^2-2\mu Mr-4M^2}{2r^4} \tilde{h}_{rr}, \\ 
Q^\sharp &=& \frac{1}{f} \tilde{h}_{tt} - f \tilde{h}_{rr}. 
\end{eqnarray*} 
The conservation (Bianchi) identities read 
\begin{eqnarray*}  
0 &=& \frac{\partial}{\partial t} Q^{tt} 
+ \frac{\partial}{\partial r} Q^{tr} 
+ \frac{2(r-M)}{r^2 f} Q^{tr} 
- \frac{\lambda}{2r^2} Q^t, \\ 
0 &=& \frac{\partial}{\partial t} Q^{tr} 
+ \frac{\partial}{\partial r} Q^{rr} 
+ \frac{Mf}{r^2} Q^{tt} 
+ \frac{2r-5M}{r^2 f} Q^{rr}
- \frac{\lambda}{2r^2} Q^r 
- \frac{f}{r} Q^\flat, \\ 
0 &=& \frac{\partial}{\partial t} Q^t 
+ \frac{\partial}{\partial r} Q^r 
+ \frac{2}{r} Q^r + Q^\flat - \frac{\mu}{2r^2} Q^\sharp. 
\end{eqnarray*}  
The Zerilli-Moncrief function is 
\[
\Psi_{\rm even} = \frac{2r}{\mu+2} \biggl[ \tilde{K}   
+ \frac{2f}{\Lambda} \biggl( f \tilde{h}_{rr} 
- r \frac{\partial}{\partial r} \tilde{K} \biggr) \biggr], 
\]
where $\Lambda = \mu + 6M/r$, and its source term is 
\begin{eqnarray*} 
S_{\rm even} &=& \frac{4}{\Lambda} Q^r - \frac{1}{r} Q^\sharp 
+ \frac{2}{(\mu+2)\Lambda} \biggl\{ 
2r^2 f \frac{\partial}{\partial r} 
     \biggl( f Q^{tt} - \frac{1}{f} Q^{rr} \biggr) 
+ \frac{24M}{\Lambda} Q^{rr} + 2r f Q^\flat 
\\ & & \mbox{} 
- \frac{r}{\Lambda} 
\biggl[ \mu(\mu-2) + 12(\mu-3) \frac{M}{r} + 84\frac{M^2}{r^2} \biggr]   
\biggl( f Q^{tt} - \frac{1}{f} Q^{rr} \biggr) \biggr\}. 
\end{eqnarray*} 

The odd-parity perturbations are $h_t$, $h_r$, and $h_2$. Under a
gauge transformation generated by $\xi$ they change according to
$h_{t} \to h'_{t} = h_{t} + \Delta h_{t}$, etc., with 
\begin{eqnarray*} 
\Delta h_{t} &=& - \frac{\partial}{\partial t} \xi, \\
\Delta h_{r} &=& - \frac{\partial}{\partial r} \xi 
+ \frac{2}{r} \xi, \\
\Delta h_2 &=& - 2\xi. 
\end{eqnarray*} 
The gauge-invariant fields are
\begin{eqnarray*} 
\tilde{h}_t &=& h_t - \frac{1}{2} \frac{\partial}{\partial t} h_2, \\ 
\tilde{h}_r &=& h_r - \frac{1}{2} \frac{\partial}{\partial r} h_2 
+ \frac{1}{r} h_2. 
\end{eqnarray*} 
The field equations are 
\begin{eqnarray*} 
P^t &=& - \frac{\partial^2}{\partial t \partial r} \tilde{h}_r 
+ \frac{\partial^2}{\partial r^2} \tilde{h}_t 
- \frac{2}{r} \frac{\partial}{\partial t} \tilde{h}_r \
- \frac{\lambda r - 4M}{r^3 f} \tilde{h}_t, \\ 
P^r &=& \frac{\partial^2}{\partial t^2} \tilde{h}_r 
- \frac{\partial^2}{\partial t \partial r} \tilde{h}_t 
+ \frac{2}{r} \frac{\partial}{\partial t} \tilde{h}_t 
+ \frac{\mu f}{r^2} \tilde{h}_r, \\ 
P &=& -\frac{1}{f} \frac{\partial}{\partial t} \tilde{h}_t 
+ f \frac{\partial}{\partial r} \tilde{h}_r 
+ \frac{2M}{r^2} \tilde{h}_r. 
\end{eqnarray*} 
The conservation (Bianchi) identity reads 
\[
0 = \frac{\partial}{\partial t} P^t 
+ \frac{\partial}{\partial r} P^r 
+ \frac{2}{r} P^r - \frac{\mu}{r^2} P. 
\]
The Cunningham-Price-Moncrief function is 
\[
\Psi_{\rm odd} = \frac{2r}{\mu} \biggl( 
\frac{\partial}{\partial r} \tilde{h}_t - 
\frac{\partial}{\partial t} \tilde{h}_r - \frac{2}{r} \tilde{h}_t 
\biggr), 
\]
and its source term is 
\[
S_{\rm odd} = \frac{2r}{\mu} \biggl( 
\frac{1}{f} \frac{\partial}{\partial t} P^r 
+ f \frac{\partial}{\partial r} P^t + \frac{2M}{r^2} P^t \biggr). 
\]

\section{Perturbation equations in $(u,r)$ coordinates} 
\label{D}

In this Appendix the two-dimensional Schwarzschild metric is written
in the form 
\[
ds^2 = -f\, du^2 - 2\, dudr, 
\]
with $f = 1-2M/r$. In these coordinates the wave operator is given by 
\[
\Box \Psi = \biggl( -2\frac{\partial^2}{\partial u \partial r} 
+ \frac{\partial}{\partial r} f \frac{\partial}{\partial r} \biggr) 
\Psi. 
\]
We set $\lambda := l(l+1) = \mu + 2$ and $\mu := (l-1)(l+2) 
= \lambda - 2$. 

The even-parity perturbations are $h_{uu}$, $h_{ur}$, $h_{rr}$, $j_u$,
$j_r$, $K$, and $G$. Under a gauge transformation generated by
$\xi_u$, $\xi_r$, and $\xi$ they change according to $h_{uu} \to
h'_{uu} = h_{uu} + \Delta h_{uu}$, etc., with   
\begin{eqnarray*} 
\Delta h_{uu} &=& -2 \frac{\partial}{\partial u} \xi_u 
- \frac{2M}{r^2} \xi_u + \frac{2Mf}{r^2} \xi_r, \\ 
\Delta h_{ur} &=& -\frac{\partial}{\partial r} \xi_u 
- \frac{\partial}{\partial u} \xi_r + \frac{2M}{r^2} \xi_r, \\
\Delta h_{rr} &=& -2 \frac{\partial}{\partial r} \xi_r, \\
\Delta j_u &=& -\frac{\partial}{\partial u} \xi - \xi_u, \\ 
\Delta j_r &=& -\frac{\partial}{\partial r} \xi - \xi_r 
+ \frac{2}{r} \xi, \\
\Delta K &=& -\frac{2f}{r} \xi_r + \frac{2}{r} \xi_u 
+ \frac{\lambda}{r^2} \xi, \\ 
\Delta G &=& -\frac{2}{r^2} \xi. 
\end{eqnarray*}     
The gauge-invariant fields are 
\begin{eqnarray*} 
\tilde{h}_{uu} &=& h_{uu} - 2 \frac{\partial}{\partial u} j_u 
- \frac{2M}{r^2} j_u + \frac{2Mf}{r^2} j_r 
+ r^2 \frac{\partial^2}{\partial u^2} G 
+ M \frac{\partial}{\partial u} G 
- Mf \frac{\partial}{\partial r} G, \\
\tilde{h}_{ur} &=& h_{ur}  
- \frac{\partial}{\partial r} j_u 
- \frac{\partial}{\partial u} j_r
+ \frac{2M}{r^2} j_r
+ r^2 \frac{\partial^2}{\partial u \partial r} G 
+ r\frac{\partial}{\partial u} G
- M\frac{\partial}{\partial r} G, \\ 
\tilde{h}_{rr} &=& h_{rr} 
- 2 \frac{\partial}{\partial r} j_r 
+ r^2 \frac{\partial^2}{\partial r^2} G
+ 2r \frac{\partial}{\partial r} G, \\
\tilde{K} &=& K + \frac{2}{r} j_u - \frac{2f}{r} j_r 
+ rf \frac{\partial}{\partial r} G - r \frac{\partial}{\partial u} G 
+ \frac{\lambda}{2} G. 
\end{eqnarray*} 
The field equations are
\begin{eqnarray*} 
Q^{uu} &=& -\frac{\partial^2}{\partial r^2} \tilde{K} 
- \frac{2}{r} \frac{\partial}{\partial r} \tilde{K} 
+ \frac{1}{r} \frac{\partial}{\partial u} \tilde{h}_{rr} 
+ \frac{f}{r} \frac{\partial}{\partial r} \tilde{h}_{rr}
- \frac{2}{r} \frac{\partial}{\partial r} \tilde{h}_{ur} 
+ \frac{\lambda r + 4M}{2r^3} \tilde{h}_{rr}, \\ 
Q^{ur} &=& \frac{\partial^2}{\partial u \partial r} \tilde{K} 
+ \frac{2}{r} \frac{\partial}{\partial u} \tilde{K} 
- \frac{r-M}{r^2} \frac{\partial}{\partial r} \tilde{K} 
- \frac{f}{r} \frac{\partial}{\partial u} \tilde{h}_{rr} 
+ \frac{1}{r} \frac{\partial}{\partial r} \tilde{h}_{uu} 
+ \frac{1}{r^2} \tilde{h}_{uu} - \frac{\lambda+4}{2r^2} \tilde{h}_{ur} 
+ \frac{f}{r^2} \tilde{h}_{rr} + \frac{\mu}{2r^2} \tilde{K}, \\ 
Q^{rr} &=& -\frac{\partial^2}{\partial u^2} \tilde{K} 
- \frac{r-M}{r^2} \frac{\partial}{\partial u} \tilde{K} 
+ \frac{(r-M)f}{r^2} \frac{\partial}{\partial r} \tilde{K}
- \frac{1}{r} \frac{\partial}{\partial u} \tilde{h}_{uu}  
+ \frac{2f}{r} \frac{\partial}{\partial u} \tilde{h}_{ur} 
- \frac{f}{r} \frac{\partial}{\partial r} \tilde{h}_{uu}  
\\ & & \mbox{}
+ \frac{\mu r + 4M}{2r^3} \tilde{h}_{uu}
+ \frac{2f}{r^2} \tilde{h}_{ur}  
- \frac{f^2}{r^2} \tilde{h}_{rr} 
- \frac{\mu f}{2r^2} \tilde{K}, \\ 
Q^{u} &=& \frac{\partial}{\partial u} \tilde{h}_{rr}
- \frac{\partial}{\partial r} \tilde{h}_{ur} 
+ \frac{\partial}{\partial r} \tilde{K} 
+ \frac{2}{r} \tilde{h}_{ur} 
- \frac{r-M}{r^2} \tilde{h}_{rr}, \\
Q^{r} &=& -\frac{\partial}{\partial u} \tilde{h}_{ur}
+ \frac{\partial}{\partial r} \tilde{h}_{uu} 
+ \frac{\partial}{\partial u} \tilde{K} 
- f \frac{\partial}{\partial r} \tilde{K} 
- \frac{2(r-M)}{r^2} \tilde{h}_{ur}
+ \frac{(r-M)f}{r^2} \tilde{h}_{rr}, \\
Q^\flat &=& -\frac{\partial^2}{\partial u^2} \tilde{h}_{rr} 
+ 2 \frac{\partial^2}{\partial u \partial r} \tilde{h}_{ur} 
- \frac{\partial^2}{\partial r^2} \tilde{h}_{uu} 
- 2 \frac{\partial^2}{\partial u \partial r} \tilde{K} 
+ f \frac{\partial^2}{\partial r^2} \tilde{K} 
+ \frac{r-M}{r^2} \frac{\partial}{\partial u} \tilde{h}_{rr} 
- \frac{2}{r} \frac{\partial}{\partial u} \tilde{K}
- \frac{2}{r} \frac{\partial}{\partial r} \tilde{h}_{uu} 
\\ & & \mbox{} 
+ \frac{2(r-M)}{r^2} \frac{\partial}{\partial r} \tilde{h}_{ur}  
- \frac{(r-M)f}{r^2} \frac{\partial}{\partial r} \tilde{h}_{rr} 
+ \frac{2(r-M)}{r^2} \frac{\partial}{\partial r} \tilde{K}  
+ \frac{\lambda}{r^2} \tilde{h}_{ur} 
- \frac{\lambda r^2-2\mu Mr-4M^2}{2r^4} \tilde{h}_{rr}, \\ 
Q^\sharp &=& 2 \tilde{h}_{ur} - f \tilde{h}_{rr}. 
\end{eqnarray*} 
The conservation (Bianchi) identities read 
\begin{eqnarray*}  
0 &=& \frac{\partial}{\partial u} Q^{uu} 
+ \frac{\partial}{\partial r} Q^{ur} 
- \frac{M}{r^2} Q^{uu} 
+ \frac{2}{r} Q^{ur} 
- \frac{\lambda}{2r^2} Q^u + \frac{1}{r} Q^\flat, \\ 
0 &=& \frac{\partial}{\partial u} Q^{ur} 
+ \frac{\partial}{\partial r} Q^{rr} 
+ \frac{Mf}{r^2} Q^{uu} 
+ \frac{2M}{r^2} Q^{ur} 
+ \frac{2}{r} Q^{rr}
- \frac{\lambda}{2r^2} Q^r 
- \frac{f}{r} Q^\flat, \\ 
0 &=& \frac{\partial}{\partial u} Q^u 
+ \frac{\partial}{\partial r} Q^r 
+ \frac{2}{r} Q^r + Q^\flat - \frac{\mu}{2r^2} Q^\sharp. 
\end{eqnarray*}  
The Zerilli-Moncrief function is 
\[
\Psi_{\rm even} = \frac{2r}{\mu+2} \biggl[ \tilde{K}   
+ \frac{2}{\Lambda} \biggl( \tilde{h}_{uu} - 2f \tilde{h}_{ur} 
+ f^2 \tilde{h}_{rr} + r \frac{\partial}{\partial u} \tilde{K}
- r f \frac{\partial}{\partial r} \tilde{K} \biggr) \biggr], 
\]
where $\Lambda = \mu + 6M/r$, and its source term is 
\begin{eqnarray*} 
S_{\rm even} &=& \frac{4}{\Lambda} Q^r - \frac{1}{r} Q^\sharp 
+ \frac{2}{(\mu+2)\Lambda} \biggl\{ 
- 2r^2 \biggl( \frac{\partial}{\partial u} 
- f\frac{\partial}{\partial r} \biggr) 
  \biggl( f Q^{uu} + 2 Q^{ur} \biggr) 
+ \frac{24M}{\Lambda} Q^{rr} + 2r f Q^\flat 
\\ & & \mbox{} 
- \frac{r}{\Lambda} 
\biggl[ \mu(\mu-2) + 12(\mu-3) \frac{M}{r} + 84\frac{M^2}{r^2} \biggr]   
\biggl( f Q^{uu} + 2 Q^{ur} \biggr) \biggr\}. 
\end{eqnarray*} 

The odd-parity perturbations are $h_u$, $h_r$, and $h_2$. Under a
gauge transformation generated by $\xi$ they change according to
$h_{u} \to h'_{u} = h_{u} + \Delta h_{u}$, etc., with 
\begin{eqnarray*} 
\Delta h_{u} &=& - \frac{\partial}{\partial u} \xi, \\
\Delta h_{r} &=& - \frac{\partial}{\partial r} \xi 
+ \frac{2}{r} \xi, \\
\Delta h_2 &=& - 2\xi. 
\end{eqnarray*} 
The gauge-invariant fields are
\begin{eqnarray*} 
\tilde{h}_u &=& h_u - \frac{1}{2} \frac{\partial}{\partial u} h_2, \\ 
\tilde{h}_r &=& h_r - \frac{1}{2} \frac{\partial}{\partial r} h_2 
+ \frac{1}{r} h_2. 
\end{eqnarray*} 
The field equations are 
\begin{eqnarray*} 
P^u &=& - \frac{\partial^2}{\partial u \partial r} \tilde{h}_r 
+ \frac{\partial^2}{\partial r^2} \tilde{h}_u 
- \frac{2}{r} \frac{\partial}{\partial u} \tilde{h}_r \
- \frac{2}{r^2} \tilde{h}_u - \frac{\mu}{r^2} \tilde{h}_r, \\ 
P^r &=& \frac{\partial^2}{\partial u^2} \tilde{h}_r 
- \frac{\partial^2}{\partial u \partial r} \tilde{h}_u 
+ \frac{2}{r} \frac{\partial}{\partial u} \tilde{h}_u 
- \frac{\mu}{r^2} \tilde{h}_u + \frac{\mu f}{r^2} \tilde{h}_r, \\  
P &=& -\frac{\partial}{\partial u} \tilde{h}_r 
- \frac{\partial}{\partial r} \tilde{h}_u 
+ f \frac{\partial}{\partial r} \tilde{h}_r 
+ \frac{2M}{r^2} \tilde{h}_r. 
\end{eqnarray*} 
The conservation (Bianchi) identity reads 
\[
0 = \frac{\partial}{\partial u} P^u 
+ \frac{\partial}{\partial r} P^r 
+ \frac{2}{r} P^r - \frac{\mu}{r^2} P. 
\]
The Cunningham-Price-Moncrief function is 
\[
\Psi_{\rm odd} = \frac{2r}{\mu} \biggl( 
\frac{\partial}{\partial r} \tilde{h}_u - 
\frac{\partial}{\partial u} \tilde{h}_r - \frac{2}{r} \tilde{h}_u 
\biggr),  
\]
and its source term is 
\[
S_{\rm odd} = \frac{2r}{\mu} \biggl( 
- \frac{\partial}{\partial u} P^u 
+ f \frac{\partial}{\partial r} P^u 
+ \frac{\partial}{\partial r} P^r + \frac{2M}{r^2} P^u \biggr). 
\]

\section{Perturbation equations in $(v,r)$ coordinates} 
\label{E}

In this Appendix the two-dimensional Schwarzschild metric is written
in the form 
\[
ds^2 = -f\, dv^2 + 2\, dvdr, 
\]
with $f = 1-2M/r$. In these coordinates the wave operator is given by 
\[
\Box \Psi = \biggl( 2\frac{\partial^2}{\partial v \partial r} 
+ \frac{\partial}{\partial r} f \frac{\partial}{\partial r} \biggr) 
\Psi. 
\]
We set $\lambda := l(l+1) = \mu + 2$ and $\mu := (l-1)(l+2) 
= \lambda - 2$. 

The even-parity perturbations are $h_{vv}$, $h_{vr}$, $h_{rr}$, $j_v$,
$j_r$, $K$, and $G$. Under a gauge transformation generated by
$\xi_v$, $\xi_r$, and $\xi$ they change according to $h_{vv} \to
h'_{vv} = h_{vv} + \Delta h_{vv}$, etc., with   
\begin{eqnarray*} 
\Delta h_{vv} &=& -2 \frac{\partial}{\partial v} \xi_v 
+ \frac{2M}{r^2} \xi_v + \frac{2Mf}{r^2} \xi_r, \\ 
\Delta h_{vr} &=& -\frac{\partial}{\partial r} \xi_v 
- \frac{\partial}{\partial v} \xi_r - \frac{2M}{r^2} \xi_r, \\
\Delta h_{rr} &=& -2 \frac{\partial}{\partial r} \xi_r, \\
\Delta j_v &=& -\frac{\partial}{\partial v} \xi - \xi_v, \\ 
\Delta j_r &=& -\frac{\partial}{\partial r} \xi - \xi_r 
+ \frac{2}{r} \xi, \\
\Delta K &=& -\frac{2f}{r} \xi_r - \frac{2}{r} \xi_v 
+ \frac{\lambda}{r^2} \xi, \\ 
\Delta G &=& -\frac{2}{r^2} \xi. 
\end{eqnarray*}     
The gauge-invariant fields are 
\begin{eqnarray*} 
\tilde{h}_{vv} &=& h_{vv} - 2 \frac{\partial}{\partial v} j_v 
+ \frac{2M}{r^2} j_v + \frac{2Mf}{r^2} j_r 
+ r^2 \frac{\partial^2}{\partial v^2} G 
- M \frac{\partial}{\partial v} G 
- Mf \frac{\partial}{\partial r} G, \\
\tilde{h}_{vr} &=& h_{vr}  
- \frac{\partial}{\partial r} j_v 
- \frac{\partial}{\partial v} j_r
- \frac{2M}{r^2} j_r
+ r^2 \frac{\partial^2}{\partial v \partial r} G 
+ r\frac{\partial}{\partial v} G
+ M\frac{\partial}{\partial r} G, \\ 
\tilde{h}_{rr} &=& h_{rr} 
- 2 \frac{\partial}{\partial r} j_r 
+ r^2 \frac{\partial^2}{\partial r^2} G
+ 2r \frac{\partial}{\partial r} G, \\
\tilde{K} &=& K - \frac{2}{r} j_v - \frac{2f}{r} j_r 
+ rf \frac{\partial}{\partial r} G + r \frac{\partial}{\partial v} G  
+ \frac{\lambda}{2} G. 
\end{eqnarray*} 
The field equations are
\begin{eqnarray*} 
Q^{vv} &=& -\frac{\partial^2}{\partial r^2} \tilde{K} 
- \frac{2}{r} \frac{\partial}{\partial r} \tilde{K} 
- \frac{1}{r} \frac{\partial}{\partial v} \tilde{h}_{rr} 
+ \frac{f}{r} \frac{\partial}{\partial r} \tilde{h}_{rr}
+ \frac{2}{r} \frac{\partial}{\partial r} \tilde{h}_{vr} 
+ \frac{\lambda r + 4M}{2r^3} \tilde{h}_{rr}, \\ 
Q^{vr} &=& \frac{\partial^2}{\partial v \partial r} \tilde{K} 
+ \frac{2}{r} \frac{\partial}{\partial v} \tilde{K} 
+ \frac{r-M}{r^2} \frac{\partial}{\partial r} \tilde{K} 
- \frac{f}{r} \frac{\partial}{\partial v} \tilde{h}_{rr} 
- \frac{1}{r} \frac{\partial}{\partial r} \tilde{h}_{vv} 
- \frac{1}{r^2} \tilde{h}_{vv} - \frac{\lambda+4}{2r^2} \tilde{h}_{vr} 
- \frac{f}{r^2} \tilde{h}_{rr} - \frac{\mu}{2r^2} \tilde{K}, \\ 
Q^{rr} &=& -\frac{\partial^2}{\partial v^2} \tilde{K} 
+ \frac{r-M}{r^2} \frac{\partial}{\partial v} \tilde{K} 
+ \frac{(r-M)f}{r^2} \frac{\partial}{\partial r} \tilde{K}
+ \frac{1}{r} \frac{\partial}{\partial v} \tilde{h}_{vv}  
+ \frac{2f}{r} \frac{\partial}{\partial v} \tilde{h}_{vr} 
- \frac{f}{r} \frac{\partial}{\partial r} \tilde{h}_{vv}  
\\ & & \mbox{}
+ \frac{\mu r + 4M}{2r^3} \tilde{h}_{vv}
- \frac{2f}{r^2} \tilde{h}_{vr}  
- \frac{f^2}{r^2} \tilde{h}_{rr} 
- \frac{\mu f}{2r^2} \tilde{K}, \\ 
Q^{v} &=& \frac{\partial}{\partial v} \tilde{h}_{rr}
- \frac{\partial}{\partial r} \tilde{h}_{vr} 
- \frac{\partial}{\partial r} \tilde{K} 
+ \frac{2}{r} \tilde{h}_{vr} 
+ \frac{r-M}{r^2} \tilde{h}_{rr}, \\
Q^{r} &=& -\frac{\partial}{\partial v} \tilde{h}_{vr}
+ \frac{\partial}{\partial r} \tilde{h}_{vv} 
- \frac{\partial}{\partial v} \tilde{K} 
- f \frac{\partial}{\partial r} \tilde{K} 
+ \frac{2(r-M)}{r^2} \tilde{h}_{vr}
+ \frac{(r-M)f}{r^2} \tilde{h}_{rr}, \\
Q^\flat &=& -\frac{\partial^2}{\partial v^2} \tilde{h}_{rr} 
+ 2 \frac{\partial^2}{\partial v \partial r} \tilde{h}_{vr} 
- \frac{\partial^2}{\partial r^2} \tilde{h}_{vv} 
+ 2 \frac{\partial^2}{\partial v \partial r} \tilde{K} 
+ f \frac{\partial^2}{\partial r^2} \tilde{K} 
- \frac{r-M}{r^2} \frac{\partial}{\partial v} \tilde{h}_{rr} 
+ \frac{2}{r} \frac{\partial}{\partial v} \tilde{K}
- \frac{2}{r} \frac{\partial}{\partial r} \tilde{h}_{vv} 
\\ & & \mbox{} 
- \frac{2(r-M)}{r^2} \frac{\partial}{\partial r} \tilde{h}_{vr}  
- \frac{(r-M)f}{r^2} \frac{\partial}{\partial r} \tilde{h}_{rr} 
+ \frac{2(r-M)}{r^2} \frac{\partial}{\partial r} \tilde{K}  
- \frac{\lambda}{r^2} \tilde{h}_{vr} 
- \frac{\lambda r^2-2\mu Mr-4M^2}{2r^4} \tilde{h}_{rr}, \\ 
Q^\sharp &=& -2 \tilde{h}_{vr} - f \tilde{h}_{rr}. 
\end{eqnarray*} 
The conservation (Bianchi) identities read 
\begin{eqnarray*}  
0 &=& \frac{\partial}{\partial v} Q^{vv} 
+ \frac{\partial}{\partial r} Q^{vr} 
+ \frac{M}{r^2} Q^{vv} 
+ \frac{2}{r} Q^{vr} 
- \frac{\lambda}{2r^2} Q^v - \frac{1}{r} Q^\flat, \\ 
0 &=& \frac{\partial}{\partial v} Q^{vr} 
+ \frac{\partial}{\partial r} Q^{rr} 
+ \frac{Mf}{r^2} Q^{vv} 
- \frac{2M}{r^2} Q^{vr} 
+ \frac{2}{r} Q^{rr}
- \frac{\lambda}{2r^2} Q^r 
- \frac{f}{r} Q^\flat, \\ 
0 &=& \frac{\partial}{\partial v} Q^v 
+ \frac{\partial}{\partial r} Q^r 
+ \frac{2}{r} Q^r + Q^\flat - \frac{\mu}{2r^2} Q^\sharp. 
\end{eqnarray*}  
The Zerilli-Moncrief function is 
\[
\Psi_{\rm even} = \frac{2r}{\mu+2} \biggl[ \tilde{K}   
+ \frac{2}{\Lambda} \biggl( \tilde{h}_{vv} + 2f \tilde{h}_{vr} 
+ f^2 \tilde{h}_{rr} - r \frac{\partial}{\partial v} \tilde{K}
- r f \frac{\partial}{\partial r} \tilde{K} \biggr) \biggr], 
\]
where $\Lambda = \mu + 6M/r$, and its source term is 
\begin{eqnarray*} 
S_{\rm even} &=& \frac{4}{\Lambda} Q^r - \frac{1}{r} Q^\sharp 
+ \frac{2}{(\mu+2)\Lambda} \biggl\{ 
2r^2 \biggl( \frac{\partial}{\partial v} 
+ f\frac{\partial}{\partial r} \biggr) 
  \biggl( f Q^{vv} - 2 Q^{vr} \biggr) 
+ \frac{24M}{\Lambda} Q^{rr} + 2r f Q^\flat 
\\ & & \mbox{} 
- \frac{r}{\Lambda} 
\biggl[ \mu(\mu-2) + 12(\mu-3) \frac{M}{r} + 84\frac{M^2}{r^2} \biggr]   
\biggl( f Q^{vv} - 2 Q^{vr} \biggr) \biggr\}. 
\end{eqnarray*} 

The odd-parity perturbations are $h_v$, $h_r$, and $h_2$. Under a
gauge transformation generated by $\xi$ they change according to
$h_{v} \to h'_{v} = h_{v} + \Delta h_{v}$, etc., with 
\begin{eqnarray*} 
\Delta h_{v} &=& - \frac{\partial}{\partial v} \xi, \\
\Delta h_{r} &=& - \frac{\partial}{\partial r} \xi 
+ \frac{2}{r} \xi, \\
\Delta h_2 &=& - 2\xi. 
\end{eqnarray*} 
The gauge-invariant fields are
\begin{eqnarray*} 
\tilde{h}_v &=& h_v - \frac{1}{2} \frac{\partial}{\partial v} h_2, \\ 
\tilde{h}_r &=& h_r - \frac{1}{2} \frac{\partial}{\partial r} h_2 
+ \frac{1}{r} h_2. 
\end{eqnarray*} 
The field equations are 
\begin{eqnarray*} 
P^v &=& - \frac{\partial^2}{\partial v \partial r} \tilde{h}_r 
+ \frac{\partial^2}{\partial r^2} \tilde{h}_v 
- \frac{2}{r} \frac{\partial}{\partial v} \tilde{h}_r \
- \frac{2}{r^2} \tilde{h}_v + \frac{\mu}{r^2} \tilde{h}_r, \\ 
P^r &=& \frac{\partial^2}{\partial v^2} \tilde{h}_r 
- \frac{\partial^2}{\partial v \partial r} \tilde{h}_v 
+ \frac{2}{r} \frac{\partial}{\partial v} \tilde{h}_v 
+ \frac{\mu}{r^2} \tilde{h}_v + \frac{\mu f}{r^2} \tilde{h}_r, \\  
P &=& \frac{\partial}{\partial v} \tilde{h}_r 
+ \frac{\partial}{\partial r} \tilde{h}_v 
+ f \frac{\partial}{\partial r} \tilde{h}_r 
+ \frac{2M}{r^2} \tilde{h}_r. 
\end{eqnarray*} 
The conservation (Bianchi) identity reads 
\[
0 = \frac{\partial}{\partial v} P^v 
+ \frac{\partial}{\partial r} P^r 
+ \frac{2}{r} P^r - \frac{\mu}{r^2} P. 
\]
The Cunningham-Price-Moncrief function is 
\[
\Psi_{\rm odd} = \frac{2r}{\mu} \biggl( 
\frac{\partial}{\partial r} \tilde{h}_v - 
\frac{\partial}{\partial v} \tilde{h}_r - \frac{2}{r} \tilde{h}_v 
\biggr),  
\]
and its source term is 
\[
S_{\rm odd} = \frac{2r}{\mu} \biggl( 
\frac{\partial}{\partial v} P^v 
+ f \frac{\partial}{\partial r} P^v 
- \frac{\partial}{\partial r} P^r + \frac{2M}{r^2} P^v \biggr). 
\]

\end{widetext} 

\bibliography{bhpert} 

\begin{thebibliography}{20}
\expandafter\ifx\csname natexlab\endcsname\relax\def\natexlab#1{#1}\fi
\expandafter\ifx\csname bibnamefont\endcsname\relax
  \def\bibnamefont#1{#1}\fi
\expandafter\ifx\csname bibfnamefont\endcsname\relax
  \def\bibfnamefont#1{#1}\fi
\expandafter\ifx\csname citenamefont\endcsname\relax
  \def\citenamefont#1{#1}\fi
\expandafter\ifx\csname url\endcsname\relax
  \def\url#1{\texttt{#1}}\fi
\expandafter\ifx\csname urlprefix\endcsname\relax\def\urlprefix{URL }\fi
\providecommand{\bibinfo}[2]{#2}
\providecommand{\eprint}[2][]{\url{#2}}

\bibitem[{\citenamefont{Regge and Wheeler}(1957)}]{regge-wheeler:57}
\bibinfo{author}{\bibfnamefont{T.}~\bibnamefont{Regge}} \bibnamefont{and}
  \bibinfo{author}{\bibfnamefont{J.~A.} \bibnamefont{Wheeler}},
  \bibinfo{journal}{Phys. Rev.} \textbf{\bibinfo{volume}{108}},
  \bibinfo{pages}{1063} (\bibinfo{year}{1957}).

\bibitem[{\citenamefont{Vishveshwara}(1970)}]{vishveshwara:70}
\bibinfo{author}{\bibfnamefont{C.~V.} \bibnamefont{Vishveshwara}},
  \bibinfo{journal}{Phys. Rev. D} \textbf{\bibinfo{volume}{1}},
  \bibinfo{pages}{2870} (\bibinfo{year}{1970}).

\bibitem[{\citenamefont{Zerilli}(1970)}]{zerilli:70}
\bibinfo{author}{\bibfnamefont{F.~J.} \bibnamefont{Zerilli}},
  \bibinfo{journal}{Phys. Rev. D} \textbf{\bibinfo{volume}{2}},
  \bibinfo{pages}{2141} (\bibinfo{year}{1970}).

\bibitem[{\citenamefont{Chandrasekhar}(1983)}]{chandrasekhar:83}
\bibinfo{author}{\bibfnamefont{S.}~\bibnamefont{Chandrasekhar}},
  \emph{\bibinfo{title}{The mathematical theory of black holes}}
  (\bibinfo{publisher}{Clarendon Press}, \bibinfo{address}{Oxford, England},
  \bibinfo{year}{1983}).

\bibitem[{\citenamefont{Frolov and Novikov}(1998)}]{frolov-novikov:98}
\bibinfo{author}{\bibfnamefont{V.~P.} \bibnamefont{Frolov}} \bibnamefont{and}
  \bibinfo{author}{\bibfnamefont{I.~D.} \bibnamefont{Novikov}},
  \emph{\bibinfo{title}{Black hole physics: Basic concepts and new
  developments}} (\bibinfo{publisher}{Kluwer}, \bibinfo{address}{Dordrecht, The
  Netherlands}, \bibinfo{year}{1998}).

\bibitem[{\citenamefont{Martel}(2004)}]{martel:04}
\bibinfo{author}{\bibfnamefont{K.}~\bibnamefont{Martel}},
  \bibinfo{journal}{Phys. Rev. D} \textbf{\bibinfo{volume}{69}},
  \bibinfo{pages}{044025} (\bibinfo{year}{2004}).

\bibitem[{\citenamefont{Sasaki and Tagoshi}(2004)}]{sasaki-tagoshi:03}
\bibinfo{author}{\bibfnamefont{M.}~\bibnamefont{Sasaki}} \bibnamefont{and}
  \bibinfo{author}{\bibfnamefont{H.}~\bibnamefont{Tagoshi}},
  \bibinfo{journal}{Living Rev. Relativity} \textbf{\bibinfo{volume}{6}}
  (\bibinfo{year}{2004}), \bibinfo{note}{6. [Online article]: cited on \today,
  http://www.livingreviews.org/lrr-2003-6}.

\bibitem[{\citenamefont{Gleiser et~al.}(2000)\citenamefont{Gleiser, Nicasio,
  Price, and Pullin}}]{gleiser-etal:00}
\bibinfo{author}{\bibfnamefont{R.}~\bibnamefont{Gleiser}},
  \bibinfo{author}{\bibfnamefont{C.}~\bibnamefont{Nicasio}},
  \bibinfo{author}{\bibfnamefont{R.}~\bibnamefont{Price}}, \bibnamefont{and}
  \bibinfo{author}{\bibfnamefont{J.}~\bibnamefont{Pullin}},
  \bibinfo{journal}{Phys. Rept.} \textbf{\bibinfo{volume}{325}},
  \bibinfo{pages}{41} (\bibinfo{year}{2000}).

\bibitem[{\citenamefont{Moncrief}(1974)}]{moncrief:74}
\bibinfo{author}{\bibfnamefont{V.}~\bibnamefont{Moncrief}},
  \bibinfo{journal}{Ann. Phys. (N.Y.)} \textbf{\bibinfo{volume}{88}},
  \bibinfo{pages}{323} (\bibinfo{year}{1974}).

\bibitem[{\citenamefont{Gerlach and Sengupta}(1980)}]{gerlach-sengupta:80}
\bibinfo{author}{\bibfnamefont{U.~H.} \bibnamefont{Gerlach}} \bibnamefont{and}
  \bibinfo{author}{\bibfnamefont{U.~K.} \bibnamefont{Sengupta}},
  \bibinfo{journal}{Phys. Rev. D} \textbf{\bibinfo{volume}{22}},
  \bibinfo{pages}{1300} (\bibinfo{year}{1980}).

\bibitem[{\citenamefont{Sarbach and Tiglio}(2001)}]{sarbach-tiglio:01}
\bibinfo{author}{\bibfnamefont{O.}~\bibnamefont{Sarbach}} \bibnamefont{and}
  \bibinfo{author}{\bibfnamefont{M.}~\bibnamefont{Tiglio}},
  \bibinfo{journal}{Phys. Rev. D} \textbf{\bibinfo{volume}{64}},
  \bibinfo{pages}{084016} (\bibinfo{year}{2001}).

\bibitem[{\citenamefont{Cunningham et~al.}(1978)\citenamefont{Cunningham,
  Price, and Moncrief}}]{cunningham-etal:78}
\bibinfo{author}{\bibfnamefont{C.~T.} \bibnamefont{Cunningham}},
  \bibinfo{author}{\bibfnamefont{R.~H.} \bibnamefont{Price}}, \bibnamefont{and}
  \bibinfo{author}{\bibfnamefont{V.}~\bibnamefont{Moncrief}},
  \bibinfo{journal}{Astrophys. J.} \textbf{\bibinfo{volume}{224}},
  \bibinfo{pages}{643} (\bibinfo{year}{1978}).

\bibitem[{\citenamefont{Jhingan and Tanaka}(2003)}]{jhingan-tanaka:03}
\bibinfo{author}{\bibfnamefont{S.}~\bibnamefont{Jhingan}} \bibnamefont{and}
  \bibinfo{author}{\bibfnamefont{T.}~\bibnamefont{Tanaka}},
  \bibinfo{journal}{Phys. Rev. D} \textbf{\bibinfo{volume}{67}},
  \bibinfo{pages}{104018} (\bibinfo{year}{2003}).

\bibitem[{\citenamefont{Martel}(2003)}]{martel:phd}
\bibinfo{author}{\bibfnamefont{K.}~\bibnamefont{Martel}}, \bibinfo{type}{Ph.d.
  thesis}, \bibinfo{school}{University of Guelph} (\bibinfo{year}{2003}).

\bibitem[{\citenamefont{Misner et~al.}(1973)\citenamefont{Misner, Thorne, and
  Wheeler}}]{MTW:73}
\bibinfo{author}{\bibfnamefont{C.~W.} \bibnamefont{Misner}},
  \bibinfo{author}{\bibfnamefont{K.~S.} \bibnamefont{Thorne}},
  \bibnamefont{and} \bibinfo{author}{\bibfnamefont{J.~A.}
  \bibnamefont{Wheeler}}, \emph{\bibinfo{title}{Gravitation}}
  (\bibinfo{publisher}{Freeman}, \bibinfo{address}{San Francisco},
  \bibinfo{year}{1973}).

\bibitem[{\citenamefont{Goldberg et~al.}(1967)\citenamefont{Goldberg,
  MacFarlane, Newman, Rohrlich, and Sudarshan}}]{goldberg-etal:67}
\bibinfo{author}{\bibfnamefont{J.}~\bibnamefont{Goldberg}},
  \bibinfo{author}{\bibfnamefont{A.}~\bibnamefont{MacFarlane}},
  \bibinfo{author}{\bibfnamefont{E.}~\bibnamefont{Newman}},
  \bibinfo{author}{\bibfnamefont{F.}~\bibnamefont{Rohrlich}}, \bibnamefont{and}
  \bibinfo{author}{\bibfnamefont{E.}~\bibnamefont{Sudarshan}},
  \bibinfo{journal}{J. Math. Phys.} \textbf{\bibinfo{volume}{8}},
  \bibinfo{pages}{2155} (\bibinfo{year}{1967}).

\bibitem[{\citenamefont{Thorne}(1980)}]{thorne:80}
\bibinfo{author}{\bibfnamefont{K.~S.} \bibnamefont{Thorne}},
  \bibinfo{journal}{Rev. Mod. Phys.} \textbf{\bibinfo{volume}{52}},
  \bibinfo{pages}{299} (\bibinfo{year}{1980}).

\bibitem[{\citenamefont{Lousto and Price}(1997)}]{lousto-price:97}
\bibinfo{author}{\bibfnamefont{C.~O.} \bibnamefont{Lousto}} \bibnamefont{and}
  \bibinfo{author}{\bibfnamefont{R.~H.} \bibnamefont{Price}},
  \bibinfo{journal}{Phys. Rev. D} \textbf{\bibinfo{volume}{55}},
  \bibinfo{pages}{2124} (\bibinfo{year}{1997}).

\bibitem[{\citenamefont{Poisson}(2004{\natexlab{a}})}]{poisson:04c}
\bibinfo{author}{\bibfnamefont{E.}~\bibnamefont{Poisson}},
  \bibinfo{journal}{Phys. Rev. D} \textbf{\bibinfo{volume}{70}},
  \bibinfo{pages}{084044} (\bibinfo{year}{2004}{\natexlab{a}}).

\bibitem[{\citenamefont{Poisson}(2004{\natexlab{b}})}]{poisson:04a}
\bibinfo{author}{\bibfnamefont{E.}~\bibnamefont{Poisson}},
  \bibinfo{journal}{Phys. Rev. D} \textbf{\bibinfo{volume}{69}},
  \bibinfo{pages}{084007} (\bibinfo{year}{2004}{\natexlab{b}}).

\end{thebibliography}
\end{document}